\documentclass[aps,prd,twocolumn,superscriptaddress,showpacs,nobibnotes]{revtex4-1}

\usepackage{graphicx,xcolor}
\usepackage{bm}
\usepackage{hyperref,subcaption}
\usepackage{amssymb,mathrsfs,mathtools,slashed}

\usepackage{cancel}
\usepackage[normalem]{ulem}

\newcommand{\als}{\alpha_\mu}
\newcommand{\alsb}{\bar\alpha_\mu}
\newcommand{\scr}[1]{\mathscr{#1}}
\newcommand{\frk}[1]{\mathfrak{#1}}
\newcommand{\ii}{{\overline{\imath}}}

\begin{document}

\title{New evidence for the rapidity evolution in Mueller--Navelet dijet production: \\ BFKL, Sudakov, and RG--invariance}

\author{A.A.~Chernyshev}\email{chernyshev@theor.jinr.ru}
\affiliation{Bogoliubov Laboratory of Theoretical Physics, Joint Institute for Nuclear Research, 141980, Dubna, Russia}

\author{M.A.~Nefedov}\email{maxim.nefedov@desy.de}
\affiliation{Physics Department, Ben-Gurion University of the Negev, 84105, Beer Sheva, Israel}

\author{V.A.~Saleev}\email{saleev.vladimir@gmail.com}
\affiliation{Bogoliubov Laboratory of Theoretical Physics, Joint Institute for Nuclear Research, 141980, Dubna, Russia}
\affiliation{Samara National Research University, 443086, Samara, Russia}

\date{\today}

\begin{abstract}
We study the effects of matching of the high-energy resummation based on the BFKL equation with initial state radiation effects, taken into account within the framework of high-energy factorization, in the production of Mueller-Navelet dijets.
We use RG-invariant solution of the NLO BFKL equation built out of eigenfunctions perturbatively constructed up to NLO to avoid the need for a special renormalization scale setting.
We demonstrate that various data sets from the FNAL Tevatron and CERN LHC can be described in this way and both, the NLL BFKL resummation and initial state radiation effects of the high-energy factorization, are crucial for the uniform description of the data across all values of the rapidity difference between the jets.
The behavior of angular distributions and ratios of angular coefficients with increasing rapidity separation between the jets provides clear evidence for the BFKL dynamics at the FNAL Tevatron and CERN LHC.
\end{abstract}

\pacs{12.38.Bx, 12.39.St, 13.85.Tp}

\maketitle

\section*{Introduction}

The Balitsky-Fadin-Kuraev-Lipatov (BFKL) equation describing the rapidity evolution in the linear regime of high-energy scattering in QCD has a history of almost $50$ years and forms a basis of the entire field of small-$x$ physics~\cite{BFKL1,BFKL2,BFKL3,BFKL4,BFKL5,BFKL6}.
The kernel of this equation is now known at leading order (LO, ${\cal O}(\als)$)~\cite{BFKL4} and next-to-LO (NLO, ${\cal O}(\als^2)$)~\cite{BFKL5}, accessing the leading logarithmic (LL $\propto \als^k Y^k$) or next-to-LL (NLL $\propto \als^{1 + k} Y^k$) resummation respectively.
The calculations of next-to-NLO corrections to the kernel are now in progress~\cite{BFKL6,Byrne:2023nqx,Fadin:2023roz,Byrne:2022wzk,Caola:2021izf,Falcioni:2021dgr, DelDuca:2021vjq,Abreu:2024xoh,Buccioni:2024gzo}.

Although the small-$x$ physics belongs among the main topics of the EIC physics program~\cite{AbdulKhalek:2021gbh}, robust phenomenological evidence for the BFKL dynamics remains elusive despite sustained efforts by the community.
Many different hard processes have been proposed to probe the BFKL dynamics at colliders~\cite{Hentschinski:22}.
One of the most popular of them is the inclusive production of a pair of jets with large transverse momenta $|{\bf p}_{\perp 1, 2}| \gg \Lambda$ and invariant mass $M^2_{1 2} \gg |{\bf p}_{\perp 1}||{\bf p}_{\perp 2}|$, such that the rapidity separation $Y \propto \ln M_{1 2}^2 / ( |{\bf p}_{\perp 1}| |{\bf p}_{\perp 2}| )$ is large: $Y \gtrsim 1/ \als$ and higher-order corrections $\propto \als^k Y^k$ have to be resummed, as suggested by Mueller and Navelet (MN)~\cite{MN:87}.

Many comprehensive studies of the MN dijet production were performed using a hybrid approach of collinear factorization (CF) with the BFKL-resummed hard scattering coefficient (HSC)~\cite{DelDuca:93,Stirling:94,Andersen:01,SabioVera:07,Szymanowski:10,Szymanowski:13,Szymanowski:14,Szymanowski:15,Szymanowski:16,Papa:13,Papa:14,Papa:15,Papa:22,Kim:23,Baldenegro:2024ndr}.
However, there are well-known difficulties with such hybrid approach associated with the appearance of large logarithmic corrections of the type $\als \ln^2 {\bf l}_\perp^2 / {\bf p}_\perp^2$ to the NLO impact-factors (IFs)~\cite{Szymanowski:10,Papa:13}, spoiling the convergence of perturbation series; the problem of these double logs is particularly important~\cite{Szymanowski:16,Altinoluk:2023hfz}.
Moreover, a lot of studies exploit the NLL Green's function, arising from the treatment of the NLO kernel being expanded over the eigenfunctions of the LO one, thus violating renormalization group (RG) invariance.

As a result, the BFKL-resummed HSC within the CF turns out to be strongly scale-sensitive, even leading to instability of calculations, which is especially evident when considering the angular distributions~\cite{Szymanowski:14,Papa:22}.
Such strong scale-sensitivity is not typical for observables in jet physics, characterized by scales exceeding tens of GeV, and most successful MN dijet phenomenology to date~\cite{Szymanowski:14,Papa:14,Papa:15,Papa:22,Kim:23} strongly relies on a particular version of Brodsky-Lepage-Mackenzie (BLM) scale-setting prescription~\cite{BLM} proposed in Ref.~\cite{BFKLP}.
In addition, those studies ignore the region $Y \sim 1 / \als$, thus missing the description of transition from fixed-order dynamics to the BFKL one.
This calls into question the robustness of those calculations as evidences for the latter.

In this paper, we resolve both of the above-mentioned issues by combining for the first time in a phenomenological study the RG-invariant solution of the NLO BKFL equation~\cite{KC:13,KC:14} with the resummation of initial state radiation (ISR) effects through the high-energy factorization (HEF) approach~\cite{HEF1,HEF2,HEF3,HEF4,HEF5,HEF6}.
The good agreement of our predictions with available experimental data, demonstrated in this paper, serves as the most precise evidence in favour of the BFKL dynamics obtained to date.

The paper is organized as follows.
In Sec.~\ref{sec:HEF}, the HEF approach is reviewed.
In Sec.~\ref{sec:BFKL}, the procedure of construction of the RG-invariant NLL Green's function is described.
In Sec.~\ref{sec:MN}, the methodology for the description of the MN dijet production is proposed and in Sec.~\ref{sec:res} the corresponding phenomenological analysis is performed.
In Sec.~\ref{sec:conc}, the conclusions are summarized.

\section{High-energy factorization} \label{sec:HEF}

It is convenient to introduce Sudakov decomposition for a four-vector $l^\mu$:
\begin{equation*}
l^\mu = \frac{l^+}{2} \, n^\mu_- + \frac{l^-}{2} \, n^\mu_+ + l^\mu_\perp,
\end{equation*}
with the basis vectors $n_\pm^\mu = \delta_0^\mu \mp \delta_3^\mu$ in the $pp$ center-of-momentum frame and $n_k^\mu = \delta_k^\mu$ for $k = 1, 2$, the corresponding derivatives: $\partial_\pm = n_\pm^\mu \partial_\mu$ and $\partial_\perp^\mu = n_k^\mu \partial^k$; also $\slashed{n} = n^\mu \gamma_\mu$.
Rapidity $y(l) = 1 / 2 \, \ln l^+ / l^-$.

\begin{figure}[t]
\centering
\includegraphics[scale=1.1]{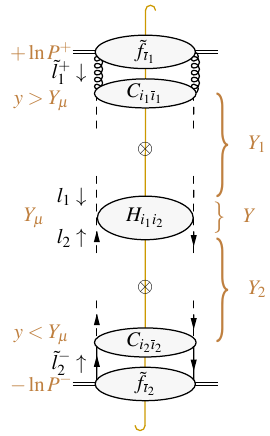}
\caption{\raggedright
Sketch of the HEF.
The ISRs are resummed by $C_{i_k \ii_k}$ and production of the dijet is described by $H_{i_1 i_2}$.
Dashed lines denote reggeized gluons and quarks.} \label{fig:HEF}
\end{figure}

The HEF approach is based on factorization properties of QCD amplitudes of production of partons highly separated in rapidity or transverse momenta.
Let's consider the auxiliary subprocess $\ii_1 (\tilde l_1) + \ii_2 (\tilde l_2) \to j_1 (p_1) + j_2 (p_2)$ with $\ii_k, j_k$ denoting gluon $g$ or quark $q$; the diagram in Fig.~\ref{fig:HEF} depicts the squared amplitude of this subprocess.
It also illustrates the factorization of the amplitude of production of particles in different rapidity regions: the interval $Y$, corresponding to the production of the dijet with rapidity $Y_\mu \propto {\cal O}(1)$, and intervals ($k = 1, 2$)
\begin{equation}
Y_k = \ln \frac{1 - z_k}{z_k} \frac{\mu_Y}{|{\bf l}_{\perp k}|},
\label{eq:Yj}
\end{equation}
containing the ISRs with rapidities $y > Y_\mu$ or $y < Y_\mu$; here $z_1 = l_1^+ / \tilde l_1^+$ and $z_2 = l_2^- / \tilde l_2^-$ with $l_1^+ = p_1^+ + p_2^+$ and $l_2^- = p_1^- + p_2^-$, rapidity scale $\mu_Y^2 \propto l_1^+ l_2^-$.
The ISR rapidities~(\ref{eq:Yj}) are the natural variables for the resummation in the HEF approach, incorporating the BFKL $\propto \ln 1 / z_k$ and Sudakov $\propto \ln \mu_Y^2 / {\bf l}_{\perp k}^2$ effects~\cite{HEF6,PRA3,MN:21}.
The resummation of the former ones is required in the regime $\mu_Y^2 \ll \tilde l_1^+ \tilde l_2^-$, i.e. $Y_k \gg 1 / \alsb$, while $Y \ll 1 / \alsb$, and of the latter ones is important even for $Y \propto 1 / \alsb$ as it has been observed in Refs.~\cite{Szymanowski:16,Altinoluk:2023hfz}, since the dijet production process is a multi-scale one.

In the Regge limit $Y_k \to \infty$, amplitude from Fig.~\ref{fig:HEF} with $n$ ISRs having $y > Y_\mu$ and $y < Y_\mu$ is factorized in rapidity space as follows~\cite{LFI:10}:
\begin{subequations} \begin{eqnarray}
\widetilde A_{\mu_{0_1} \ldots \mu_{n_1} \mu \mu_{n_2} \ldots \mu_{0_2}}^{b_{0_1} \, \ldots b_{n_1} \, b \, \, b_{n_2} \ldots b_{0_2}} =
& B_{\mu_{0_1} \ldots \mu_{n_1} -}^{b_{0_1} \ldots b_{n_1} a_{2 n_1}}(\tilde l_1, - l_1) \, D_{a_{2 n_1} a_1}(l_{\perp 1})
\nonumber \\
\times \ A_\mu^{a_1 b a_2} \times
& D_{a_2 a_{2 n_2}}(l_{\perp 2}) \, B_{+ \mu_{n_2} \ldots \mu_{0_2}}^{a_{2 n_2} b_{n_2} \ldots b_{0_2}}(- l_2, \tilde l_2)
\nonumber \\ &
+ {\cal O}\left( e^{- Y_{1, 2}}, \als^3 Y_{1, 2} \right),
\label{eq:amp}
\end{eqnarray}
amplitude $A$ describes production of the dijet via subprocess $i_1 \, (l_1) + i_2 \, (l_2) \to j_1 \, (p_1) + j_2 \, (p_2)$ with $i_k$ denoting reggeized gluon $R$ or quark $Q$, and factors
\begin{align}
\hspace{-.1cm}
B_{\mu_0 \ldots \mu_n \mp}^{b_0 \ldots b_n a_{2 n}}(\tilde l, - l) & =
\gamma_{\mu_0 \mp}^{b_0 a_0}(- \tilde l)
\prod_{k = 1}^n \theta( \pm ( y(\tilde l_k - \tilde l_{k + 1}) - Y_\mu ) )
\nonumber \\ & \hspace{-1.5cm} \times
D_{a_{2 k - 2} a_{2 k - 1}}(\tilde l_{\perp k}) \,
\Gamma_{+ \mu_k -}^{a_{2 k - 1} b_k a_{2 k}}( \tilde l_k, - \tilde l_{k + 1} )
\label{eq:amp2}
\end{align} \end{subequations}
absorb the ISRs; here $b_k$ and $\mu_k$ are the color and kinematical index of the $k^{\rm th.}$ ISR, $a_k$ is the color index of the reggeon and $\pm$ stands for its large light-cone component.
Each block in Eqs.~(\ref{eq:amp}) and~(\ref{eq:amp2}) is constructed of reggeon propagators $D(l_\perp)$ and {\it gauge-invariant} effective vertices $\gamma_{\mu \pm}$ and $\Gamma_{+ \mu -}$, satisfying Slavnov–Taylor identities.
Factorization in Eq.~(\ref{eq:amp}) holds up to ${\cal O}(e^{- Y_{1, 2}})$ and NNLL approximation (${\cal O}(\als^3 Y_{1, 2})$), at which multi-reggeon exchange effects start to contribute, spoiling the factorization~\cite{DelDuca:2001gu,Caron-Huot:2013fea,BFKL6}.

Following from Eq.~(\ref{eq:amp}), cross section for the dijet production within the HEF approach can be represented as follows~\cite{HEF1,HEF2,HEF3,HEF4,HEF5,HEF6}:
\begin{eqnarray}
\sigma & = &
\int \frac{d x_1}{x_1} \frac{1}{\pi} \int\limits_{{\bf l}_{\perp 1}}
\Phi_{i_1}\left( x_1, {\bf l}_{\perp 1}^2, \mu^2 \right)
\nonumber \\ & \times &
\int \frac{d x_2}{x_2} \frac{1}{\pi} \int\limits_{{\bf l}_{\perp 2}}
\Phi_{i_2}\left( x_2, {\bf l}_{\perp 2}^2, \mu^2 \right)
H_{i_1 i_2}\left( x_{1, 2}, {\bf l}_{\perp 1, 2}, \als \right)
\nonumber \\ && \hspace{2.5cm}
+ \ {\cal O}\left( \Lambda^\# / \mu^\#, e^{- Y_{1, 2}}, \als^3 \, Y_{1, 2} \right).
\label{eq:HEF}
\end{eqnarray}
Process-dependent HSC $H_{i_1 i_2}$ describe production of the dijet, while process-independent unintegrated PDFs (UPDFs) $\Phi_{i_k}$ are responsible for the resummation of the ISR effects.
Factorization in Eq.~(\ref{eq:HEF}) holds in the leading-twist approximation and has the same type of the corrections as Eq.~(\ref{eq:amp}).

The HSC in Eq.~(\ref{eq:HEF}) is expressed in terms of the amplitude $A$ with reggeized initial-state partons $i_{1}$ and $i_2$ in Eq.~(\ref{eq:amp}) in a standard way~\cite{HEF6,PRA3} and is given by:
\begin{eqnarray}
\frac{d H_{i_1 i_2}}{d^2 {\bf p}_{\perp 1} d y_1 d^2 {\bf p}_{\perp 2} d y_2} & = &
\nonumber \\ && \hspace{-2.cm}
\delta^{(4)}\left( l_1 + l_2 - p_1 - p_2 \right)
\frac{\overline{\mid A_{i_1 i_2} \mid^2}}{32 \pi^2 l_1^+ l_2^-}.
\label{eq:HSC}
\end{eqnarray}
Following from Eq.~(\ref{eq:amp}), for the external reggeized gluons the factor $l_k^\pm / (2 |{\bf l}_{\perp k}|)$ should be included into the amplitude, while the prescription for spinors of external reggeized quarks is $u(l_{\parallel k})$, averaging over spinor and color indices proceeds as usual.
To compute the amputated amplitudes in $A_{i_1 i_2}$, we use gauge-invariant effective action approach by L.~Lipatov~\cite{EFT1,EFT2}, which we describe below.

The QCD in the Regge limit is described by the effective action~\cite{EFT1,EFT2}:
\begin{equation}
S = \int d^4 x \ {\scr L}(x),
\hspace{1.cm}
{\scr L}(x) = {\scr L}_0(x) + {\scr L}_{\rm I}(x),
\label{eq:EA}
\end{equation}
operating with several degrees of freedom, these are gluon vector fields $A_\mu(x) = A_\mu^a(x) \, T_a$ and spinor quark fields $Q(x)$, transforming under local group in a standard way, and reggeized gluon scalar fields ${\frk A}_\pm(x) = {\frk A}_\pm^a(x) \, T_a$ and reggeized quark spinor fields ${\frk Q}_\pm(x)$, which are {\it invariant} w.r.t. local gauge transformations.
The latter two are subjects of the kinematical constraints: $\partial_\mp {\frk A}_\pm(x) = 0$ and $\partial_\mp {\frk Q}_\pm(x) = 0$, following from Eq.~(\ref{eq:amp}); in addition, $\slashed{n}_\mp {\frk Q}_\pm(x) = 0$, canceling a structures with the small light-cone components.

The Lagrangian~(\ref{eq:EA}) contains the standard QCD one:
\begin{equation*}
{\scr L}(x) \supset
- \frac{1}{2} \, F^2(x) + i \overline Q(x) \, \slashed{\scr D} \, Q(x),
\end{equation*}
where $F_{\mu_1 \mu_2}(x) = - i / g \left[ {\scr D}_{\mu_1}, {\scr D}_{\mu_2} \right]$ is field-strength tensor and ${\scr D}_\mu = \partial_\mu + i g A_\mu(x)$ is covariant derivative; one should add the corresponding gauge-fixing term to this Lagrangian.
We use the Feynman gauge for internal propagators and define polarization vectors for the final-state gluons in axial gauge $n_+\varepsilon(p)=0$.
Feynman rules, following from this Lagrangian, are well known~\cite{LFI:10}.

The free Lagrangian~(\ref{eq:EA}) includes forms, quadratic and non-diagonal in the reggeized gluon fields:
\begin{subequations} \begin{equation}
{\scr L}_0(x) \supset
4 {\rm tr} \ {\frk A}_+(x) \, \partial_\perp^2 \, {\frk A}_-(x),
\label{eq:L0A}
\end{equation}
and the reggeized quark fields:
\begin{equation}
{\scr L}_0(x) \supset
i \overline {\frk Q}_+(x) \, \slashed{\partial}_\perp \, {\frk Q}_-(x) + {\rm h.c.},
\label{eq:L0Q}
\end{equation} \end{subequations}
which describe propagation of the reggeized partons in cross channel.
The interaction Lagrangian~(\ref{eq:EA}) includes non-local terms, describing interactions of the partons and reggeized partons, one has for the reggeized gluons:
\begin{subequations} \begin{eqnarray}
\hspace{-.7cm}
{\scr L}_{\rm I}(x) & \supset &
\frac{i}{g} \, {\rm tr}\left[
{\frk A}_+(x) \, \partial_\perp^2 \, \partial_- \left( [- \infty, x_+] - [x_+, - \infty] \right) \right]
\label{eq:LIA} \\ & + &
\frac{i}{g} \, {\rm tr}\left[
{\frk A}_-(x) \, \partial_\perp^2 \, \partial_+ \left( [- \infty, x_-] - [x_-, - \infty] \right) \right],
\nonumber
\end{eqnarray}
and the reggeized quarks:
\begin{eqnarray}
{\scr L}_{\rm I}(x) & \supset &
\frac{i}{2} \, \overline {\frk Q}_+(x) \, \slashed{\partial}
\left( [x_+, - \infty] + [x_+, + \infty] \right) Q(x)
\label{eq:LIQ} \\
& + &
\frac{i}{2} \, \overline {\frk Q}_-(x) \, \slashed{\partial}
\left( [x_-, - \infty] + [x_-, + \infty] \right) Q(x)
+ {\rm h.c.}; \nonumber
\end{eqnarray} \end{subequations}
semi-infinite straight gauge links are defined as follows:
\begin{equation}
[\pm \infty, x_\mp] = {\rm P} \,
{\rm exp}\left[ - \frac{i g}{2} \int\limits_{\pm \infty}^{x_\mp} d x_\mp' \ A_\pm(x_\pm, x_\mp', x_\perp) \right],
\label{eq:gl}
\end{equation}
here ${\rm P}$ denote path-ordering and $A_\pm(x) = A_\mu(x) \, n_\pm^\mu$.
In Eqs.~(\ref{eq:LIA}) and~(\ref{eq:LIQ}), the Hermitian version of the effective action~\cite{BZ:18} is used, which is necessary for loop computations~\cite{MN:19}.
Expansion of Eq.~(\ref{eq:gl}) in perturbation series generates an infinite number of the induced terms, restoring the gauge-invariance order-by-order in coupling constant~\cite{EFT1,EFT2,EFT3}; however, at every fixed-order one need only a few of them.

Free Lagrangian in Eq.~(\ref{eq:L0A}) leads to the propagator of the reggeized gluon: $- i \delta^{a_1 a_2} / \left( 2 l_\perp^2 \right)$; in the case of Eq.~(\ref{eq:L0Q}), it is convenient~\cite{EFT2} to remove ${\cal O}(g^0)$ term in Eq.~(\ref{eq:LIQ}) by shifting $Q(x) \to Q(x) + \hat P_+ \, {\frk Q}_+(x) + \hat P_- \, {\frk Q}_-(x)$ with $\hat P_\pm = (1 / 4) \, \slashed{n}_\mp \slashed{n}_\pm$ being a projector, so the propagator of the reggeized quark: $i \hat P_\pm / \slashed{l}_\perp$.
Up to ${\cal O}({g^3})$, Eq.~(\ref{eq:gl}) gives:
\begin{equation*}
\text{r.h.s.~(\ref{eq:gl})} \supset I
- i g \, \partial_\pm^{-1} A_\pm(x) - g^2 \, \partial_\pm^{-1} A_\pm(x) \, \partial_\pm^{-1} A_\pm(x);
\end{equation*}
using this expansion, from Eq.~(\ref{eq:LIA}) one obtain the induced vertices for the reggeized gluons:
\begin{align*}
\Delta^{a b_1}_{\pm \mu_1}(l, p_1) & =
- i \delta^{a b_1} n_{\mu_1}^\mp \, l_\perp^2,
\\
\Delta^{a b_1 b_2}_{\pm \mu_1 \mu_2}(l, p_1, p_2) & =
i g \, \widetilde T^{a b_1 b_2} \, n_{\mu_1}^\mp n_{\mu_2}^\mp \, \frac{l_\perp^2}{p_1^\mp},
\\
\Delta^{a b_1 b_2 b_3}_{\pm \mu_1 \mu_2 \mu_3}(l, p_1, p_2, p_3) & =
i g^2
\left[ \frac{\widetilde T^{a b_1 b} \widetilde T_b^{\ b_2 b_3}}{p_1^\mp} + \frac{\widetilde T^{a b_2 b} \widetilde T_b^{\ b_1 b_3}}{p_2^\mp} \right]
\\ & \times
n_{\mu_1}^\mp n_{\mu_2}^\mp n_{\mu_3}^\mp \, \frac{l_\perp^2}{p_3^\mp},
\end{align*}
here $\widetilde T^{a b_1 b_2}$ are the $SU(N_c)$ generators in the adjoint representation,
and from Eq.~(\ref{eq:LIQ}) for the reggeized quarks:
\begin{align*}
\Delta^{\ b_1}_{\pm \mu_1}(l, p_1) & =
i g \, T^{b_1} \left[ \gamma_{\mu_1} - n_{\mu_1}^\mp \, \frac{\slashed{l}}{p_1^\mp} \right],
\\
\Delta^{\ b_1 b_2}_{\pm \mu_1 \mu_2}(l, p_1, p_2) & =
i g^2 \left[ \frac{T^{b_1} T^{b_2}}{p_1^\mp} + \frac{T^{b_2} T^{b_1}}{p_2^\mp} \right]
\\ & \times
n_{\mu_1}^\mp n_{\mu_2}^\mp \, \frac{\slashed{l}}{p_1^\mp + p_2^\mp},
\end{align*}
here $T^b$ are the $SU(N_c)$ generators in the fundamental representation; all momenta are incoming.
With the help of these Feynman rules, one can construct the effective vertices according to the procedure of Ref.~\cite{EFT3} and compute the HSC~(\ref{eq:HSC}) in a manifestly gauge-invariant way for all possible LO subprocesses for the dijet production.
This represents the current state-of-the-art of the fixed-order HEF computations; the scheme of NLO HEF calculations has recently been proposed~\cite{HN:25}.

The HSC defined above has an on-shell-limit property:
\begin{equation*}
\int\limits_0^{2 \pi} \frac{d \phi_1 d \phi_2}{(2 \pi)^2} \lim_{{\bf l}_{\perp 1, 2}^2 \to 0}
H_{i_1 i_2}\left( x_{1, 2}, {\bf l}_{\perp 1, 2} \right) =
\hat\sigma_{\ii_1 \ii_2}\left( x_{1, 2} \right)
\end{equation*}
with $\hat\sigma_{\ii_1 \ii_2}$ being the corresponding CF HSC with, $\phi_k$ is the azimuthal angle of ${\bf l}_{\perp k}$.
This universal property has been verified for all LO HSC contributing to the dijet production in Ref.~\cite{PRA1}.

In the HEF approach, the UPDFs are perturbatively expressed in terms of collinear PDFs as follows, see Fig.~\ref{fig:HEF}:
\begin{equation}
\Phi_i(x, {\bf l}_\perp^2, \mu_Y^2) =
\sum_\ii \int\limits_x^1 \frac{d z}{z} \
C_{i \ii}\left( x, z, {\bf l}_\perp^2, \mu_Y^2, \mu_F^2 \right)
\tilde f_\ii\left( \frac{x}{z}, \mu_F^2 \right),
\label{eq:UPDF}
\end{equation}
where $\ii = g , q$ and $\tilde f_\ii(x, \mu_F^2) = x f_\ii(x, \mu_F^2)$.
Several approaches to compute process-independent resummation factors $C_{i \ii}$ exist in the HEF framework, differing by the perturbative accuracy.
The simplest possible approximation is the doubly-logarithmic one (HEF-DLA or {\it Collins-Ellis-Bl\"umlein})~\cite{HEF6,Blumlein:1995eu}, resumming only the effects $\propto \alsb^n \ln^n 1 / z \, \ln^n {\bf l}_\perp^2 / \mu_F^2$.
Despite its simplicity, this approximation is appropriate for the class of the hard inclusive~\cite{Lansberg:2021vie,Lansberg:2023kzf} and exclusive~\cite{Flett:2024htj} processes for which the effects of $\ln 1 / z$ are leading.
However, as it was already discussed above, in the case of the dijet production it is also necessary to include the resummation of the LL $\propto \als^k \ln^k {\bf l}_\perp^2 / \mu_F^2 \, \ln^k {\bf l}_\perp^2 / \mu_Y^2$ and NLL $\propto \als^k \ln^k {\bf l}_\perp^2 / \mu_{Y, F}^2$ Sudakov effects.
The HEF approach with the ISR rapidity~(\ref{eq:Yj}) incorporates both, the DLA and LL/NLL Sudakov effects~\cite{MN:21}.

As a practical tool, combining the HEF-DLA and NLL Sudakov effects, we apply Kimber-Martin-Ryskin-Watt (KMRW) formula~\cite{KMR,MRW} improved within the parton reggeization approach~\cite{PRA3}:
\begin{eqnarray}
C_{i \ii}(x, z, {\bf l}_\perp^2, \mu^2) & = &
\frac{\alpha_\mu}{2 \pi} \frac{T_i(x, {\bf l}_\perp^2, \mu^2)}{{\bf l}_\perp^2} \
\nonumber \\ & \times &
\theta\left( \Delta(|{\bf l}_\perp|, \mu) - z \right) z p_{i \ii}^{(0)}(z),
\label{eq:C}
\end{eqnarray}
where $\mu^2 \sim \mu_Y^2 \sim \mu_F^2$, $p_{i \ii}^{(0)}$ is LO DGLAP splitting functions, rapidity cutoff in accordance with the evolution variable~(\ref{eq:Yj}) is imposed by $\Delta(|{\bf l}_\perp|, \mu) = \mu / \left( |{\bf l}_\perp| + \mu \right)$, and:
\begin{equation}
- \ln T_i(x, {\bf l}_\perp^2, \mu^2) =
\frac{\alpha_\mu}{2 \pi}
\int\limits_{{\bf l}_\perp^2}^{\mu^2} \frac{d {\bf l}_\perp'^2}{{\bf l}_\perp'^2}
\big[ \tau_i({\bf l}_\perp'^2, \mu^2) + \Delta\tau_i(x, {\bf l}_\perp'^2, \mu^2) \big],
\label{eq:T}
\end{equation}
where
\begin{eqnarray*}
\tau_i({\bf l}_\perp'^2, \mu^2) & = &
\sum_\ii \int\limits_0^1 d z \
\theta\left( {\Delta({\bf l}_\perp'^2, \mu^2)} - z \right) z p_{\ii i}^{(0)}(z),
\\
\Delta\tau_i(x, {\bf l}_\perp'^2, \mu^2) & = &
\sum_\ii \int\limits_0^1 d z \
\theta\left( z - {\Delta({\bf l}_\perp'^2, \mu^2)} \right)
\\ & \times &
\bigg[ z p_{\ii i}^{(0)}(z)
- \frac{\tilde f_\ii\left( \frac{x}{z}, {\bf l}_\perp^2 \right)}{\tilde f_i(x, {\bf l}_\perp^2)} \,
\theta(z - x) \, p_{i \ii}^{(0)}(z) \bigg].
\nonumber
\end{eqnarray*}
One important property of Eq.~(\ref{eq:T}) is the correspondence with the transverse momentum dependent (TMD) factorization~\cite{CSS} up to NLL.
Expanding Eq.~(\ref{eq:T}) in the limit $|{\bf l}_\perp'| / \mu \ll 1$ over the powers of $1 - \Delta({\bf l}_\perp'^2, \mu^2) = {\cal O}\left( |{\bf l}_\perp'| / \mu \right)$, one gets:
\begin{equation}
- \ln T_i({\bf l}_T^2, \mu^2) \simeq
\frac{\alpha_\mu}{2 \pi}
\int\limits_{{\bf l}_T^2}^{\mu^2} \frac{d {\bf l}_T'^2}{{\bf l}_T'^2}
\left[ \Gamma_i^{(0)} \ln\frac{{\bf l}_T'^2}{\mu^2} + \gamma_i^{(0)} \right],
\label{eq:SFF}
\end{equation}
where $\gamma_i^{(0)} = \sum\limits_\ii \gamma_{\ii i}^{(0)}$ with $\gamma_{\ii i}^{(0)}$ being one-loop anomalous dimension and $\Gamma_i^{(0)}$ stands for one-loop cusp anomalous dimension.
Thus, Eq.~(\ref{eq:SFF}) reproduce TMD Sudakov form-factor, resumming the LL and NLL Sudakov effects.

Another important property of Eq.~(\ref{eq:C}) is exact normalization, holding for arbitrary $x$:
\begin{equation*}
\int\limits_0^{\mu^2} d {\bf l}_\perp^2 \ C_{i \ii}(x, z, {\bf l}_\perp^2, \mu^2) =
\delta_{i \ii} \, \delta(1 - z),
\end{equation*}
and, due to Eq.~(\ref{eq:UPDF}), implying the correspondence between the UPDF and PDF.

\section{BFKL Green's function} \label{sec:BFKL}

In the momentum representation, the BFKL equation for the Green's function $G$ reads:
\begin{equation}
\partial_Y G \left( {\bf l}_{\perp 1}, {\bf l}_{\perp 2}, Y \right) =
\int\limits_{{\bf l}_\perp'}  K \left( {\bf l}_{\perp 1}, {\bf l}_\perp' \right)
G \left( {\bf l}_\perp', {\bf l}_{\perp 2}, Y \right)
\label{eq:BFKL}
\end{equation}
with $K$ being the kernel.
At the LO, the kernel is conformally invariant w.r.t. $SL(2, \mathbb{C})$, so its eigenfunctions $H_{n, \gamma}^{(0)}$ are simply the conformal powers (see Eq.~(\ref{eq:EFLO})), forming the complete orthonormal basis, and the solution of Eq.~(\ref{eq:BFKL}) can be constructed as expansion over this basis.
At the NLO, the kernel is no longer conformally invariant due to running coupling effects and its complete basis of orthogonal eigenfunctions $H_{n, \gamma}$ must be constructed order-by-order in $\alsb$, ensuring the RG-invariance as proposed in Refs.~\cite{KC:13,KC:14}:
\begin{equation}
\int\limits_{{\bf l}_\perp'}  K \left( {\bf l}_\perp, {\bf l}_\perp' \right)
H_{n, \gamma}({\bf l}_\perp', \mu_R^2) =
\alsb \, \chi(n, \gamma) \, H_{n, \gamma}({\bf l}_\perp, \mu_R^2),
\label{eq:K}
\end{equation}
here $\gamma = 1 / 2 + {\rm Im} \, \gamma$ and $n \in \mathbb{Z}$ are the conformal weight and spin respectively.
In Eq.~(\ref{eq:K}), the corresponding NLO characteristic function is given by:
\begin{subequations} \begin{eqnarray}
\chi(n, \gamma) = \chi^{(0)}(n, \gamma) + \alsb \, \chi^{(1)}(n, \gamma),
\label{eq:EV}
\end{eqnarray}
where using result of Ref.~\cite{KL:00}:
\begin{eqnarray}
\chi^{(0)}(n, \gamma) & = &
2 \psi(1) - 2 {\rm Re} \, \psi\left( \gamma + \frac{n}{2} \right),
\label{eq:EVLO} \\
\chi^{(1)}(n, \gamma) & = &
- \frac{1}{2} \, b_0 \, ( \chi^{(0)}(n, \gamma) )^2
+ \Gamma_g^{(1)} \, \chi^{(0)}(n, \gamma)
\nonumber \\ && \hspace{-2.cm}
+ \frac{1}{4} \, \partial_\gamma^2 \chi^{(0)}(n, \gamma)
+ \frac{3}{2} \, \zeta(3)
- \frac{\pi^2 \cos(\pi \gamma)}{4 (1 - 2 \gamma) \sin^2(\pi \gamma)}
\nonumber \\ && \hspace{-2.cm}
\times \bigg{[} \left( 3 + \left( 1 + \frac{N_{\rm f}}{N_c^3} \right) \frac{2 + 3 \gamma (1 - \gamma)}
{(1 + 2 \gamma) (3 - 2 \gamma)} \right) \delta_{0, n}
\nonumber \\ && \hspace{-.95cm} -
\left( 1 + \frac{N_{\rm f}}{N_c^3} \right) \frac{\gamma (1 - \gamma)}
{2 (1 + 2 \gamma) (3 - 2 \gamma)} \, \delta_{2, n} \bigg{]}
\nonumber \\ && \hspace{-2.cm}
- \frac{1}{2} \left[ F(n, \gamma) + \{ \gamma \to 1 - \gamma \} \right],
\label{eq:EVNLO}
\end{eqnarray} \end{subequations}
where $b_0 = 1 / (4 \pi) \left( 11 / 3 \, C_A - 4 / 3 \, T_F N_{\rm f} \right)$ is one-loop $\beta$-function,  $\Gamma_g^{(1)} = 1 / 4 \left( 67 / 9 - 20 / 9 \, T_F N_{\rm f} / N_c - 2 \zeta(2) \right)$ is two-loop cusp anomalous dimension, definition of the function $F$ can be found in Ref.~\cite{KL:00}.
Using Eqs.~(\ref{eq:EVLO})--(\ref{eq:EVNLO}), one can verify the symmetry of Eq.~(\ref{eq:EV}) w.r.t. $\gamma \leftrightarrow 1 - \gamma$, which is necessary to preserve the projectile-target symmetry (${\bf l}_{\perp 1} \leftrightarrow {\bf l}_{\perp 2}$) of the Green's function in Eq.~(\ref{eq:BFKL}).
Following the technique of Refs.~\cite{KC:13,KC:14}, the NLO eigenfunctions are constructed as follows:
\begin{subequations} \begin{eqnarray}
H_{n, \gamma}({\bf l}_\perp, \mu_R^2) =
H^{(0)}_{n, \gamma}({\bf l}_\perp) + \alsb \,
H^{(1)}_{n, \gamma}({\bf l}_\perp, \mu_R^2),
\label{eq:EF}
\end{eqnarray}
where
\begin{eqnarray}
H^{(0)}_{n, \gamma}({\bf l}_\perp) & = & \frac{1}{\sqrt{\pi}} \, {\bf l}_\perp^{2 (\gamma - 1)} e^{i n \psi_l},
\label{eq:EFLO} \\
H^{(1)}_{n, \gamma}({\bf l}_\perp, \mu_R^2) & = &
\sum\limits_{k = 1}^\infty \, c_k(n, \gamma) \ln^k \frac{{\bf l}_\perp^2}{\mu_R^2} \,
H^{(0)}_{n, \gamma}({\bf l}_\perp),
\label{eq:EFNLO}
\end{eqnarray} \end{subequations}
there are only two non-vanishing coefficients:
\begin{equation*}
c_1(n, \gamma) = \partial_\gamma c_2 (n, \gamma),
\quad
c_2(n, \gamma) = \frac{1}{2} \, b_0 \, \frac{1}{\partial_\gamma \ln\chi^{(0)}(n, \gamma)},
\end{equation*}
and $c_{k \geq 3} = 0$.
Using Eqs.~(\ref{eq:EFLO})--(\ref{eq:EFNLO}), one can verify the completeness and orthogonality of the basis constructed of Eq.~(\ref{eq:EF}).

The RG-invariant NLL Green's function of Eq.~(\ref{eq:BFKL}) is given by:
\begin{eqnarray}
G \left( {\bf l}_{\perp 1}, {\bf l}_{\perp 2}, Y \right) & = &
\sum\limits_n \int\limits \frac{d \gamma}{2 \pi i} \
\exp\left[ \alsb \chi(n, \gamma) \, Y \right]
\nonumber \\ & \times &
H_{n, \gamma}({\bf l}_{\perp 1}, \mu_R^2) \, H_{n, \gamma}^\star({\bf l}_{\perp 2}, \mu_R^2),
\label{eq:G}
\end{eqnarray}
initial condition $G \left( {\bf l}_{\perp 1}, {\bf l}_{\perp 2}, 0 \right) = \delta^{(2)}\left( {\bf l}_{\perp 1} - {\bf l}_{\perp 2} \right)$ is satisfied due to completeness; the running coupling and NLO eigenfunctions are taken at some renormalization scale $\mu_R$.
The integral along the contour ${\rm Re} \, \gamma = 1 / 2$ in Eq.~(\ref{eq:G}) should be understood in the sense of principal value, treating the eigenfunctions as distributions due to the pole of $c_2$ at $\gamma = 1 / 2$.
Using Eqs.~(\ref{eq:EF}) and~(\ref{eq:EFNLO}), the product of the NLO eigenfunctions can be written as:
\begin{eqnarray}
H_{n, \gamma}({\bf l}_{\perp 1}, \mu_R^2) \, H_{n, \gamma}^\star({\bf l}_{\perp 2}, \mu_R^2) =
H_{n, \gamma}^{(0)}({\bf l}_{\perp 1}) \, H_{n, \gamma}^{(0) \star}({\bf l}_{\perp 2})
\nonumber \\ && \hspace{-8.cm} \times
\left[ 1 + \alsb \ln \frac{{\bf l}_{\perp 1}^2 {\bf l}_{\perp 2}^2}{\mu_R^4} \left[
\ln \frac{{\bf l}_{\perp 1}^2}{{\bf l}_{\perp 2}^2} + \partial_\gamma
\right] c_2(n, \gamma) \right],
\label{eq:Grhs1}
\end{eqnarray}
where $\Delta\psi = \psi_{l_1} - \psi_{l_2}$; substituting Eq.~(\ref{eq:Grhs1}) into Eq.~(\ref{eq:G}) and integrating by parts yields:
\begin{eqnarray}
\text{r.h.s.~(\ref{eq:G})} =
\sum\limits_n \int\limits \frac{d \gamma}{2 \pi^2 i  |{\bf l}_{\perp 1}| |{\bf l}_{\perp 2}|} \,
\exp\left[ \alsb \, \chi(n, \gamma) \, Y \right]
\label{eq:Grhs2} \\ && \hspace{-8.5cm}
H_{n, \gamma}^{(0)}({\bf l}_{\perp 1}) \, H_{n, \gamma}^{(0) \star}({\bf l}_{\perp 2})
\left[ 1 - \alsb^2 \, b_0 \, \chi^{(0)}(n, \gamma) \, Y
\ln\frac{|{\bf l}_{\perp 1}| |{\bf l}_{\perp 2}|}{\mu_R^2} \right],
\nonumber
\end{eqnarray}
one can note the cancellation of the pole at $\gamma = 1 / 2$ at this step, which takes place in Eq.~(\ref{eq:Grhs1}).
We have checked the consistency of the solution~(\ref{eq:G}) in a form~(\ref{eq:Grhs2}) with the running coupling solution found in Ref.~\cite{AG:13} and verified the agreement up to NLO, as one would expect.
We will present the details of this comparison elsewhere.

\begin{figure*}
\centering
\includegraphics[scale=0.35]{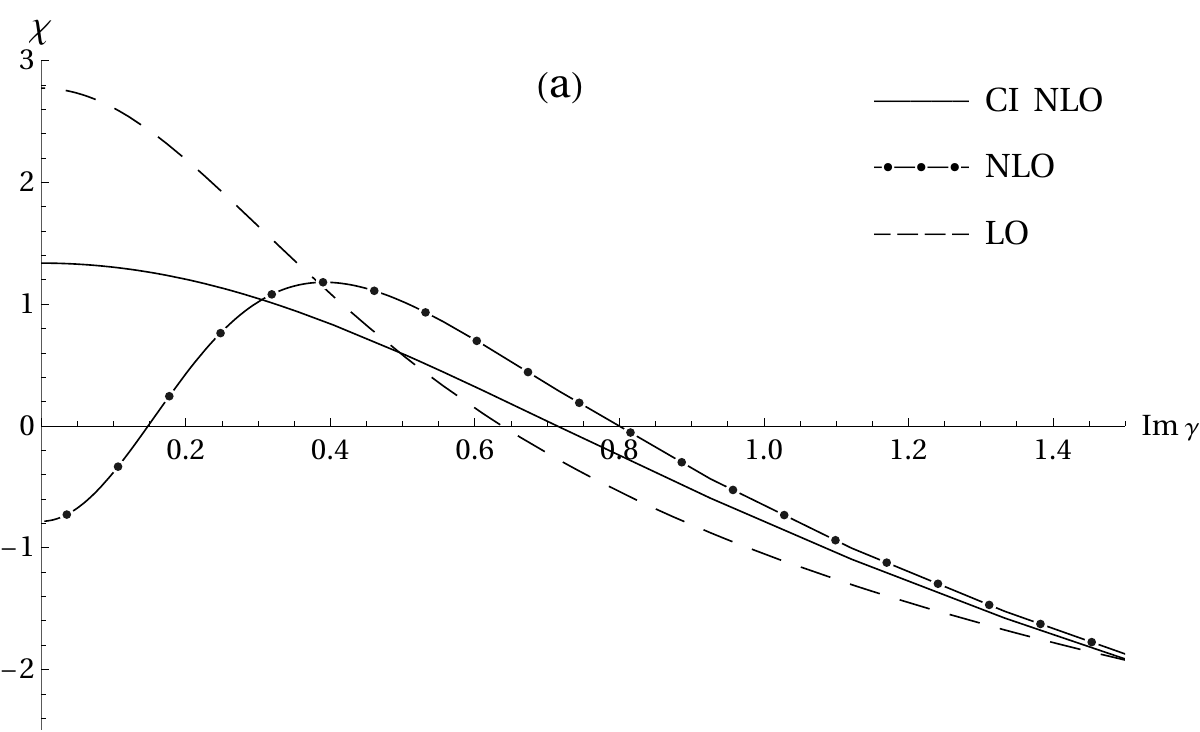}
\hspace{2.cm}
\includegraphics[scale=0.35]{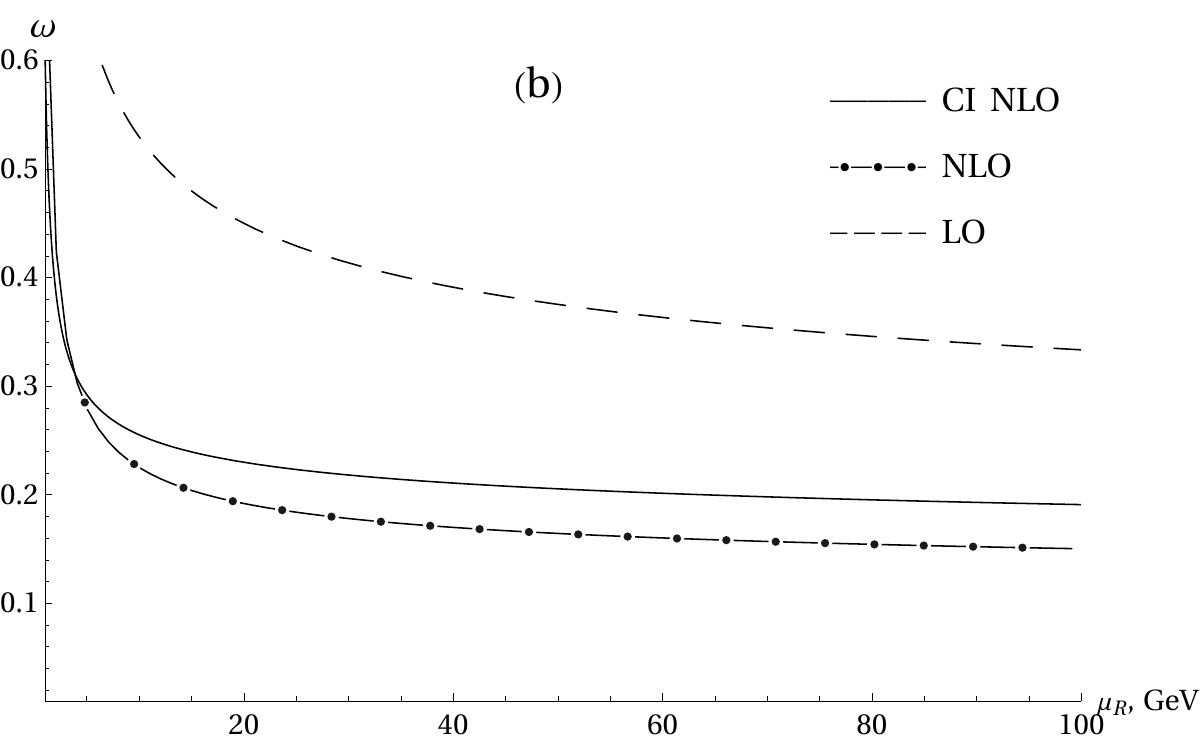}
\caption{\raggedright
The (a) characteristic function at $\text{Re}\,\gamma=1/2$ as function of $\text{Im}\,\gamma$ for $n = 0$ and $\alsb = 0.2$ and (b) intercept as function of $\mu_R$, the both for $N_c = 3$ and $N_{\rm f} = 5$.
Different curves correspond to: LO (dashed), NLO (dot-dashed), and CI NLO (solid).}\label{figA:1}
\end{figure*}

As it has been first observed in Ref.~\cite{Salam:98}, the numerically largest NLO contribution to the characteristic function~(\ref{eq:EV}) comes from \textit{(anti-) collinear poles} located at $\gamma = - n / 2$ ($1 + n / 2$).
In the vicinity of the (anti-)collinear poles the characteristic function behaves as:
\begin{eqnarray}
\chi(n, \gamma) =
\sum_{k_1 = 0}^1 \sum_{k_2 = 1}^{2 k_1 + 1}
\alsb^{k_1} \, u_{k_1, k_2}(n) \, d^{- k_2}_0(n, \gamma)
\nonumber \\ +
{\cal O}(\gamma) + \{ \gamma \to 1 - \gamma \},
\label{eq:EVp}
\end{eqnarray}
where $d_k(n, \gamma) = k + \gamma + n / 2$ and the coefficients $u_{k_1, k_2}(n)$ being defined from the pole structure of Eq.~(\ref{eq:EV}).
To study the effect of the leading and sub-leading (anti-) collinear poles in Eq.~(\ref{eq:Grhs2}), it is convenient~\cite{Salam:98,Ciafaloni:2003rd,SabioVera:05} to introduce an additional Mellin transform:
\begin{eqnarray}
\text{r.h.s.~(\ref{eq:Grhs2})} =
\sum\limits_n \int\limits \frac{d \gamma \, d M}{(2 \pi i )^2 \pi |{\bf l}_{\perp 1}| |{\bf l}_{\perp 2}|}
\frac{\exp\left[ M Y \right]}{M - \alsb \, \chi(n, \gamma)}
\nonumber \\ && \hspace{-8.5cm}
e^{i n \Delta\psi}
\left( \frac{{\bf l}_{\perp 1}^2}{{\bf l}_{\perp 2}^2} \right)^{\gamma - 1 / 2}
\left[ 1 - \alsb^2 \, b_0 \, \chi^{(0)}(n, \gamma) \, Y
\ln\frac{|{\bf l}_{\perp 1}| |{\bf l}_{\perp 2}|}{\mu_R^2} \right],
\nonumber \\
\label{eq:Grhs3}
\end{eqnarray}
here we made use of Eq.~(\ref{eq:EFLO}).
In Eq.~(\ref{eq:Grhs3}), the effects of changing the rapidity factorization scheme $Y = \ln s / s_0$ from symmetric $s_0 = |{\bf l}_{\perp 1}| |{\bf l}_{\perp 2}|$ with $|{\bf l}_{\perp 1}| \sim |{\bf l}_{\perp 2}|$ to the ``$\pm$''--scheme $s_0 = {\bf l}_{\perp 1, 2}^2$ with $|{\bf l}_{\perp 1, 2}| \gg |{\bf l}_{\perp 2, 1}|$ can be incorporated by shifting the poles $\gamma \to \gamma \mp M / 2$.
Thus, the expansion of the shifted poles in Eq.~(\ref{eq:EVp}) will contain the effects of the leading $\propto \alsb^k / \gamma^{2 k + 1}$ and sub-leading $\propto \alsb^k / \gamma^{k + 1}$ poles, so they can be resummed at all orders in $\alsb$~\cite{Salam:98,SabioVera:05,SabioVera:07} using the ``all-poles'' approximation:
\begin{eqnarray}
\Delta\chi(n, \gamma) =
\frac{1}{\alsb}
\sum_{k = 0}^\infty \bigg{[}
u_{1, 2}(n) \, \alsb - d_k(n, \gamma)
\label{eq:CIEV} \\ && \hspace{-7.cm} +
\sqrt{2 \, \alsb \left( u_{0, 1}(n) + u_{1, 1}(n) \, \alsb \right)
+ \left( u_{1, 2}(n) \, \alsb - d_k(n, \gamma) \right)^2}
\nonumber \\ && \hspace{-7.cm} -
\sum_{k_1 = 0}^1 \sum_{k_2 = 1}^{2 k_1 + 1} \,
u_{k_1, k_2}(n) \, d_{k_2}^{- k}(n, \gamma) \, \alsb^{k_1 + 1} \bigg{]}
+ \{ \gamma \to 1 - \gamma \},
\nonumber
\end{eqnarray}
where the fixed-order terms present in Eq.~(\ref{eq:EVp}) are subtracted to avoid double-counting.
Note that adding the term~(\ref{eq:CIEV}) to the characteristic function~(\ref{eq:EV}) is consistent with Eq.~(\ref{eq:K}) since $\Delta\chi$ is ${\cal O}(\alsb^2)$.
In the present paper, we refer to the addition of Eq.~(\ref{eq:CIEV}) to Eq.~(\ref{eq:EV}) as collinear improvement (CI).

The pomeron intercept, characterizing the behavior of Eq.~(\ref{eq:EV}) at $Y \to \infty$, is defined at the saddle point $(n_*, \gamma_*)$ of the appropriate characteristic function in Eq.~(\ref{eq:EV}): $\omega = \alsb \, \chi(n_*, \gamma_*)$.
At the LO, the latter one has only one saddle point, located at $\gamma_* = 1 / 2$ with the largest value at $n_* = 0$, which splits into two complex conjugate saddle points at the NLO, whose position depends on $\alsb$, see Fig.~\ref{figA:1}; however, near $\gamma_*$ up to ${\cal O}\left( (\gamma - \gamma_*)^2 \right)$:
\begin{eqnarray*}
\chi^{(0)}(0, \gamma) & \simeq &
4 \ln 2
\\
\chi^{(1)}(0, \gamma) & \simeq &
4 \ln 2 \, \Gamma_g^{(1)} - 8 \ln^2 2 \, \frac{\pi}{N_c} \, b_0 - \frac{11}{2} \, \zeta(3)
\nonumber \\ & - &
\frac{\pi^3}{128}\left( 59 + 11 \, \frac{N_{\rm f}}{N_c^3} \right) - F\left( 0, \frac{1}{2} \right),
\end{eqnarray*}
here $F(0, 1 / 2) \simeq - 3.64$.
Our results for the characteristic function and intercept are consistent with ones presented in the recent Ref.~\cite{Polizzi:2025edm}.
The CI NLO characteristic function has only one saddle point $\gamma_*$, preserving the LO definition of the intercept, see Fig.~\ref{figA:1}:
\begin{eqnarray*}
\omega & \simeq &
2.77 \, \alsb
- \left[ 17.91 - 0.13 \, \frac{N_{\rm f}}{N_c} - 2.66 \, \frac{N_{\rm f}}{N_c^3} \right] \alsb^2
\\ && \hspace{5.cm} + \Delta\chi\left( 0, \frac{1}{2} \right) \alsb;
\end{eqnarray*}
both, the NLO and CI NLO intercepts, feature a reduced scale-dependence at relatively large $\mu_R$ typical for the MN dijet production in comparison with the LO intercept and are much smaller than the LO intercept.
Note that corrections $\propto N_{\rm f} / N_c$ are relatively small~\cite{BFKL5}.
Several plots of the Green's function~(\ref{eq:Grhs3}) are presented in Appendix~\ref{ap:GF}.

\begin{figure*}[t]
\centering
\includegraphics[scale=.8]{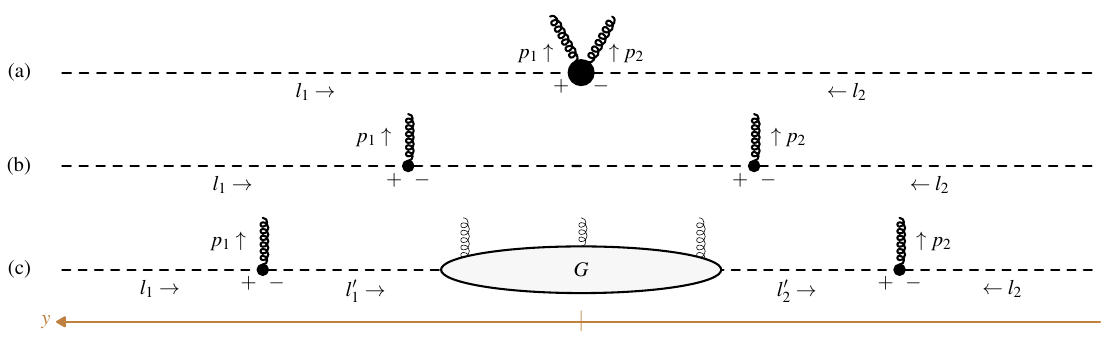}
\caption{Localization in rapidity of the different HSCs in Eq.~(\ref{eq:ms}).
Dashed lines denotes reggeized gluons and solid circles denote effective vertices.} \label{figA:0}
\end{figure*}

\section{Dijet production} \label{sec:MN}

Introducing the variable $\bar Y \coloneq \alsb \, Y$, one can distinguish two regimes for the HSC~(\ref{eq:HSC}).
They are illustrated by rapidity-space diagrams in Fig.~\ref{figA:0}.

When $\bar Y \ll 1$, the effects of the BFKL resummation of Eq.~(\ref{eq:BFKL}) are negligible and the HSC in Eq.~(\ref{eq:HSC}) can be calculated at the fixed order in $\als$ (``{\it the HEF contribution}''), see Fig.~\ref{figA:0}~(a).
The HEF HSC $H_{i_1 i_2}^{\rm (HEF)}$ for complete set of the LO amplitudes, localized in the rapidity, i.e. without internal reggeized propagators, for the production of dijet previously was computed in Sec.~II of Ref.~\cite{PRA1}.

When $\bar Y \propto 1$, the BFKL resummation should be taken into account {\it (``the BFKL-improved HEF contribution'')}, see Fig.~\ref{figA:0}~(c).
In this case, the HSC~(\ref{eq:HSC}) for the dijet production is factorized as follows~\cite{LFI:10,KL:12}:
\begin{eqnarray}
H^{\rm (BFKL)}_{i_1 i_2}  & = &
\int\limits_{{\bf l}_{\perp 1}}
I^+_{i_1} \left( {\bf l}_{\perp 1}, {\bf l}_{\perp 1}', y_1 \right)
\nonumber \\ & \times &
\int\limits_{{\bf l}_{\perp 2}}
I^-_{i_2} \left( {\bf l}_{\perp 2}, {\bf l}_{\perp 2}', y_2 \right)
G \left( {\bf l}_{\perp 1}', {\bf l}_{\perp 2}', Y \right),
\label{eq:HSC2}
\end{eqnarray}
where
\begin{eqnarray}
\frac{d I_i^\pm\left( {\bf l}_\perp, {\bf l}_\perp', y \right)}{d^2 {\bf p}_\perp \, d y} & = &
\alpha_\mu \frac{C_i}{\sqrt{2}} \frac{1}{{\bf p}_\perp^2}
\label{eq:HEFIF} \\ & \times & \nonumber
\delta(1 - l^\pm / p^\pm) \ \delta^{(2)}({\bf l}_\perp \mp {\bf l}_\perp' - {\bf p}_\perp)
\end{eqnarray}
are the HEF IFs, describing production of the jets with ${\bf p}_{\perp 1, 2} = {\bf l}_{\perp 1, 2} \mp {\bf l}_{\perp 1, 2}'$ and rapidities $y_{1, 2}$, here $C_{R / Q} = C_{A / F}$.
Let us note that the Sudakov form-factor~(\ref{eq:T}) can be moved from Eq.~(\ref{eq:C}) into the HEF IF~(\ref{eq:HEFIF}), thus, by virtue of Eq.~(\ref{eq:SFF}), the LL and NLL Sudakov logs of ${\bf l}_\perp^2 / \mu^2 \propto {\bf l}_\perp^2 / {\bf p}_\perp^2$ are resummed into $T_i\left( x, {\bf l}_\perp^2, \mu^2 \right)\times I_i^\pm\left( {\bf l}_\perp, {\bf l}_\perp', y \right)$.
In this sense, despite the expression for the HEF IFs~(\ref{eq:HEFIF}) being LO in $\als$, we go beyond the accuracy of complete NLO computations of Refs.~\cite{Szymanowski:10,Szymanowski:13,Szymanowski:14,Papa:13}, containing only terms $\propto \als^2 \ln^2\left( ({\bf l}_\perp' - {\bf p}_\perp)^2 / {\bf p}_\perp^2 \right)$ in the CF NLO IFs.
Thus, unlike the CF IFs, the HEF IFs should not receive large NLO corrections~\cite{Szymanowski:16,Altinoluk:2023hfz} and combination of the two resummation, HEF and BFKL, allows one to take into account the most important part of the radiative corrections to the LL BFKL picture of MN dijet production.

Finally, one can match the two contributions, HEF (applicable at $\bar Y \ll 1$) and BFKL-improved HEF (for $\bar Y \propto 1$), using the following subtractive matching scheme, first used in Ref.~\cite{HKNS:19}:
\begin{equation}
H_{i_1 i_2}^{\rm (HEF+BFKL)} =
H_{i_1 i_2}^{\rm (HEF)} +
H_{i_1 i_2}^{\rm (BFKL)} -
H_{i_1 i_2}^{\rm (BFKL, 0)},
\label{eq:ms}
\end{equation}
where $H_{i_1 i_2}^{\rm (BFKL, 0)}$ is the BFKL HSC in Eq.~(\ref{eq:HSC2}), but with the Green's function in Eq.~(\ref{eq:G}) replaced by its initial condition $\delta^{(2)}\left( {\bf l}_{\perp 1}' - {\bf l}_{\perp 2}' \right)$, see Fig.~\ref{figA:0}~(b).
This BFKL${}^{(0)}$-term plays two roles: at $Y \to \infty$ it coincides with the (Regge) asymptotics of $H_{i_1 i_2}^{\rm (HEF)}$ and cancels it to avoid the double-counting with the BFKL-resummation term.
For $\bar Y \sim 1$ the BFKL${}^{(0)}$-term cancels the $H_{i_1 i_2}^{\rm (BFKL)}$, since the $\delta^{(2)}\left( {\bf l}_{\perp 1}' - {\bf l}_{\perp 2}' \right)$ initial condition is already included into the Green's function~(\ref{eq:G}).
In this way, the matching scheme eliminates double counting between the HEF and BFKL-improved HEF HSCs in Eq.~(\ref{eq:ms}) and smoothly interpolates predictions between the two regimes.
The corresponding cross sections are calculated by substituting HSCs from Eq.~(\ref{eq:ms}) into Eq.~(\ref{eq:HEF}).

\section{Results} \label{sec:res}

In our computation we set $\mu = \mu_R = 2^{\xi} \sqrt{|{\bf p}_{\perp 1}| |{\bf p}_{\perp 2}|}$ with $\xi = 0$ for the central curves and $\xi = \pm 1$ to estimate the scale uncertainty of our predictions.
The numerical integration had been performed using {\tt CUBA} library~\cite{CUBA}.
We have used the improved KMRW UPDFs from Ref.~\cite{PRA3} based on parameterization of the {\tt MSTW-2008} PDFs~\cite{MSTW}.

\begin{figure*}
\centering
\begin{subfigure}[t]{0.32\textwidth}
    \centering
    \includegraphics[scale=0.327]{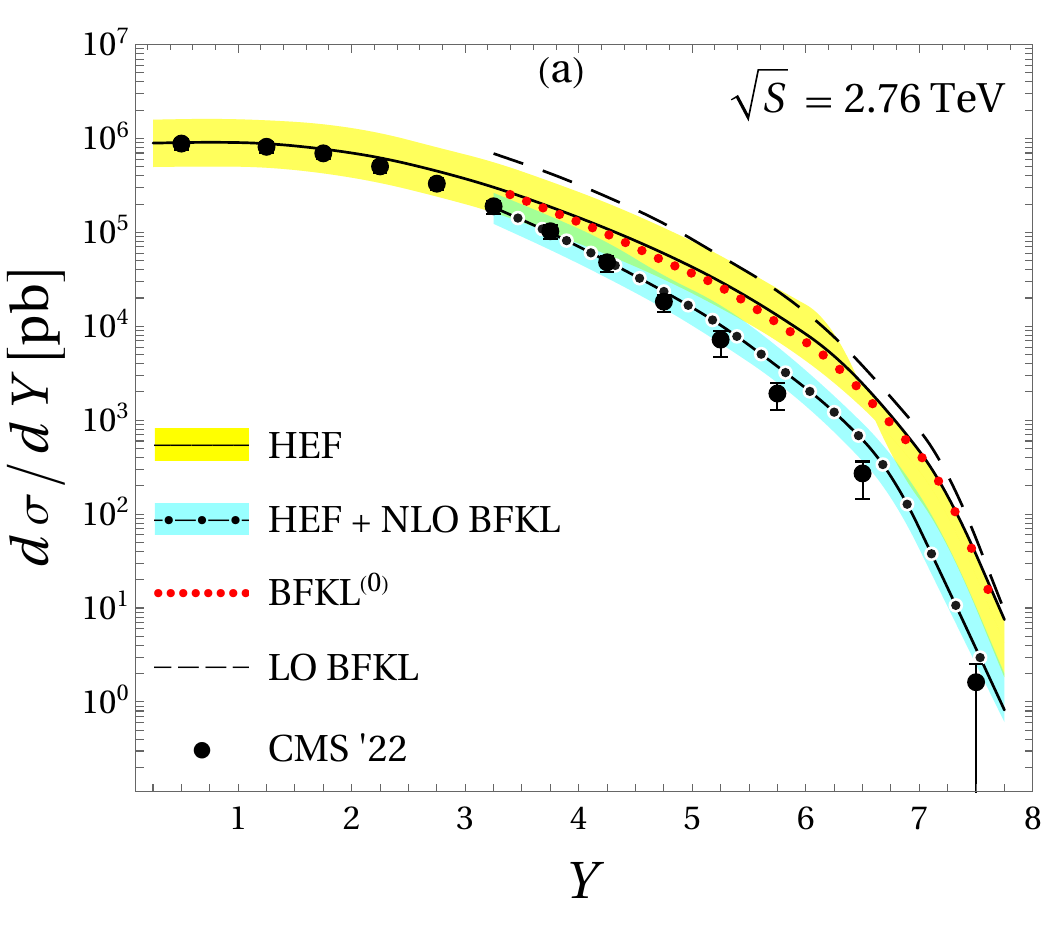}
\end{subfigure}
\begin{subfigure}[t]{0.32\textwidth}
    \centering
    \includegraphics[scale=0.33]{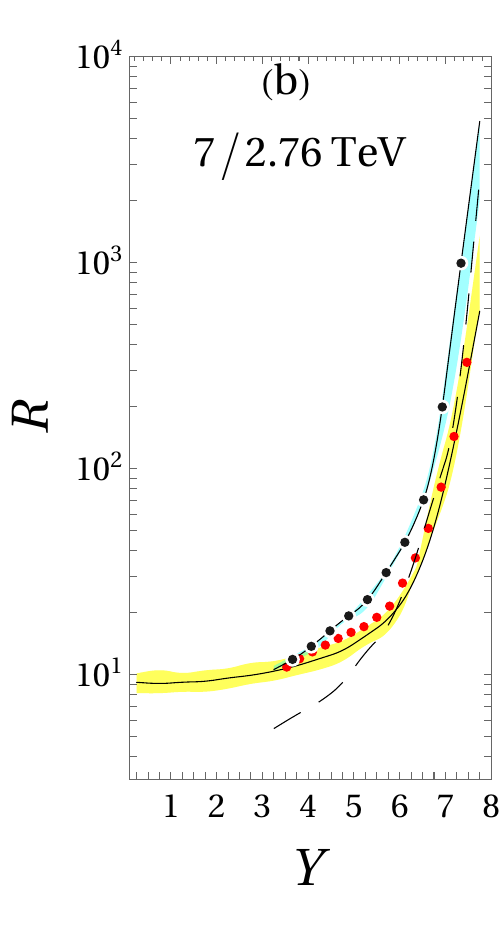}
    \includegraphics[scale=0.33]{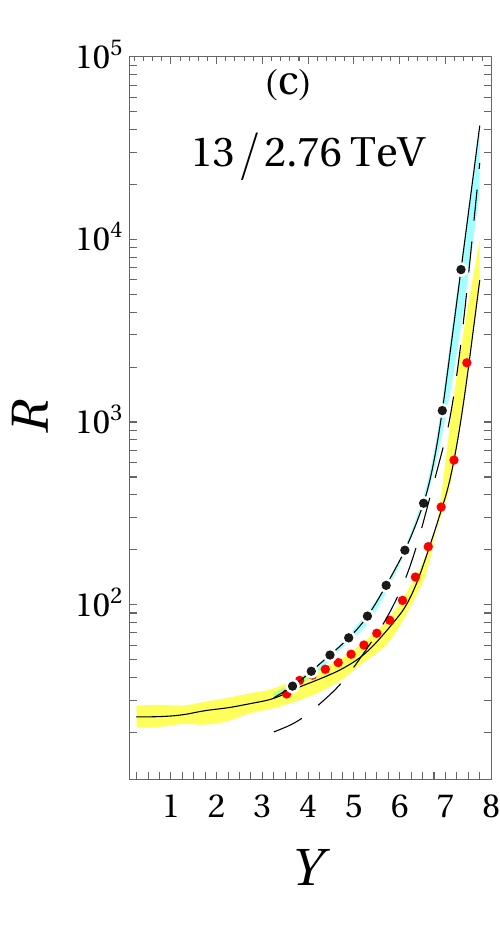}
\end{subfigure}
\begin{subfigure}{0.32\textwidth}
    \raggedleft
    \includegraphics[scale=0.33]{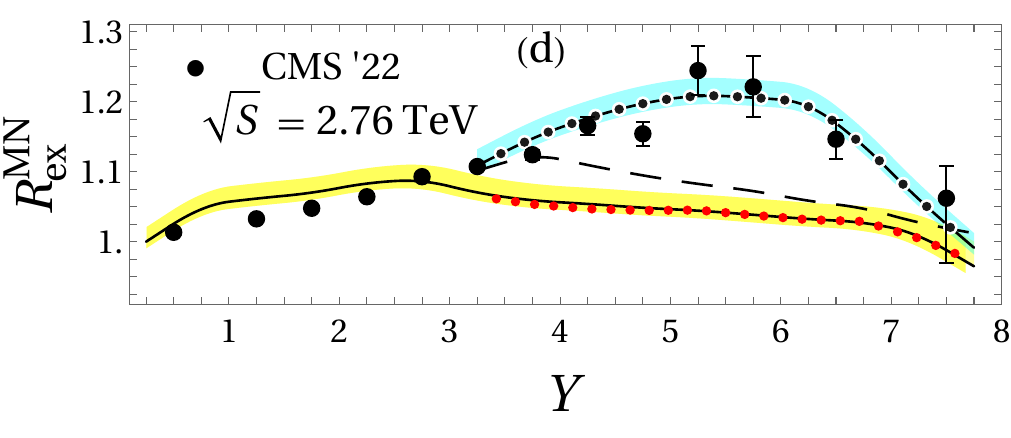}
    \vspace{0.15cm}
    \includegraphics[scale=0.33]{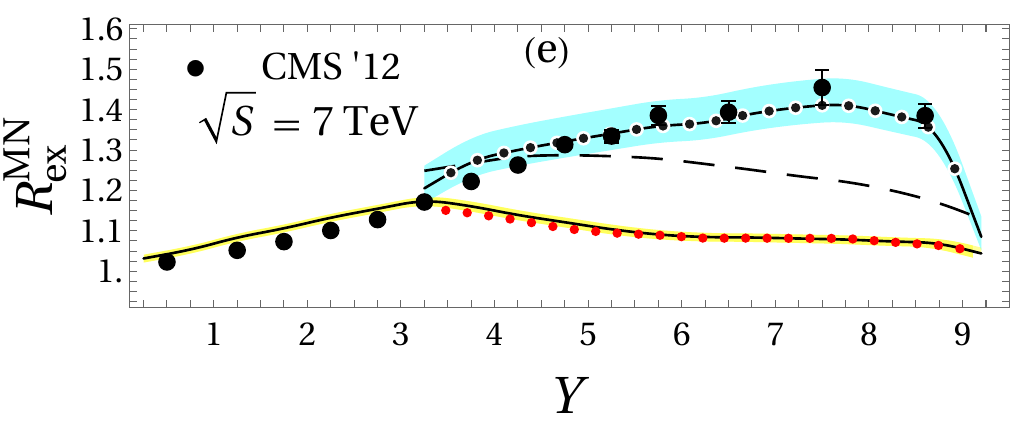}
\end{subfigure}
\\
\begin{subfigure}[t]{1.\textwidth}
    \centering
    \includegraphics[scale=0.327]{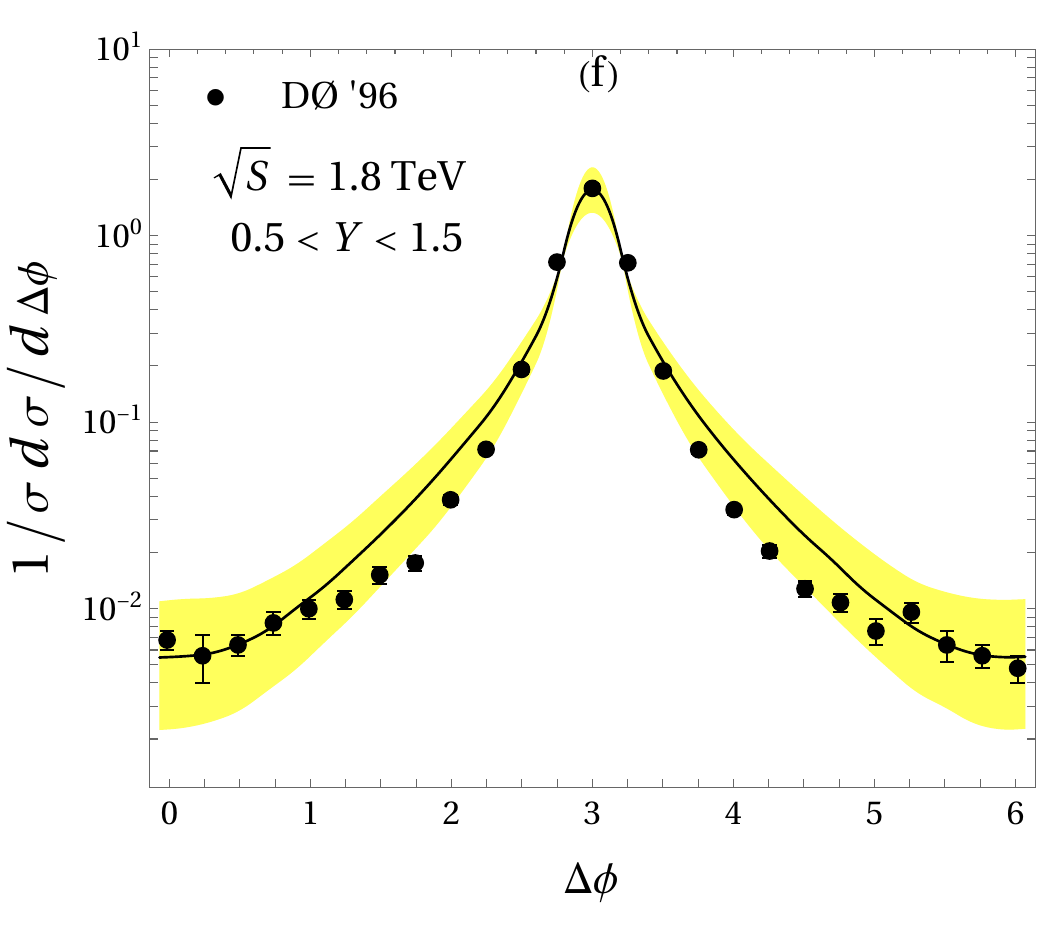}
    \includegraphics[scale=0.327]{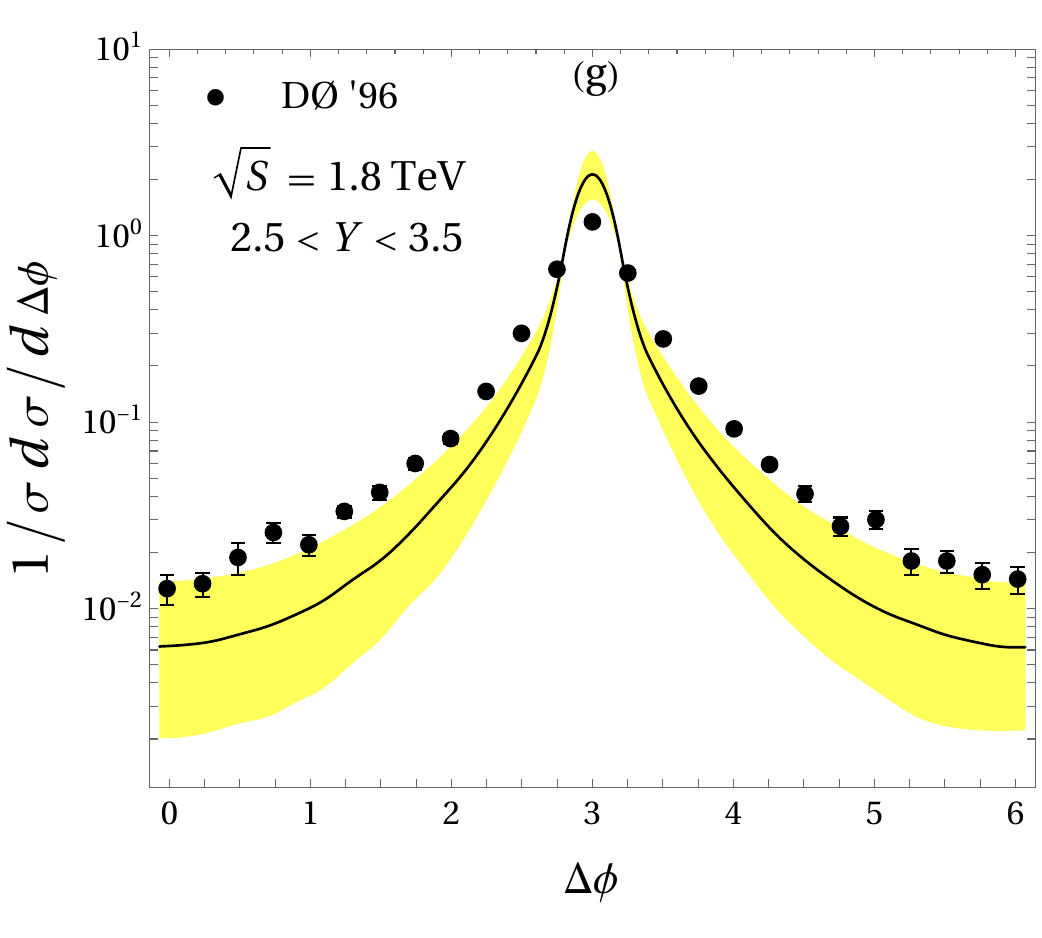}
    \includegraphics[scale=0.327]{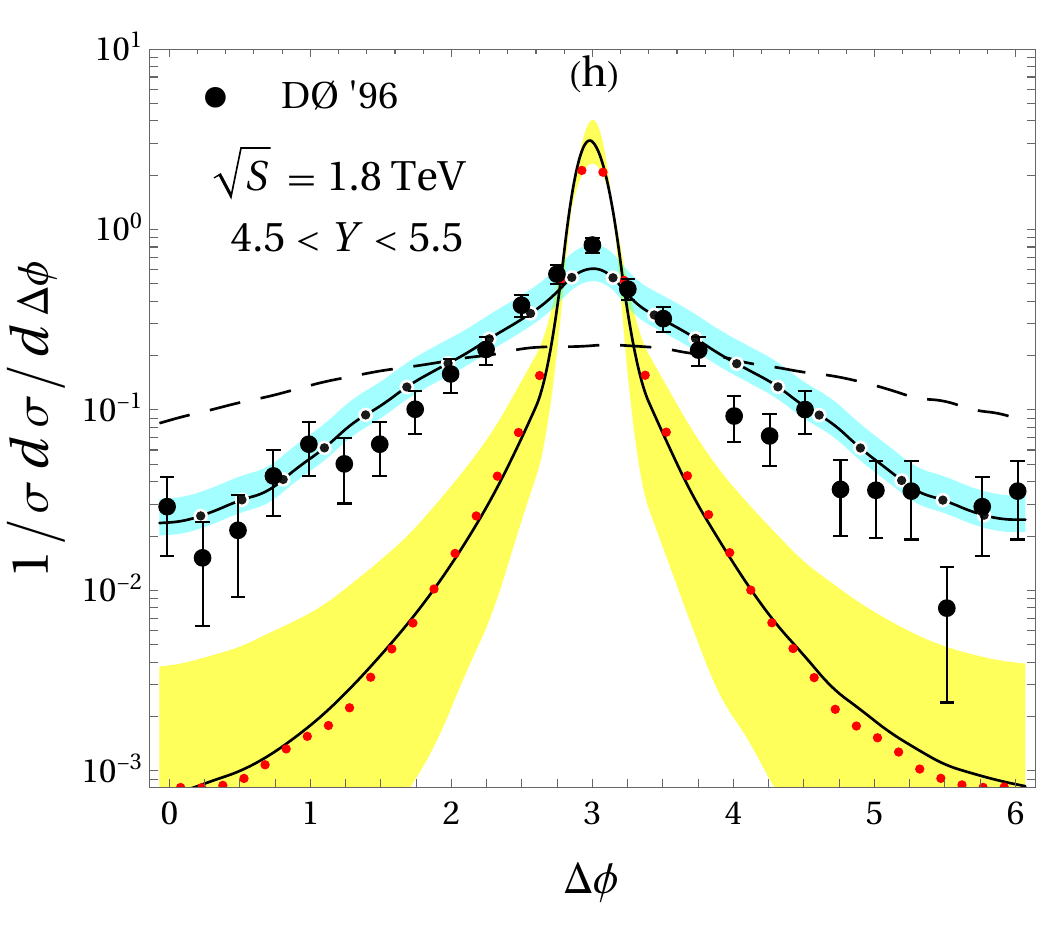}
\end{subfigure}
\\
\begin{subfigure}[t]{1.\textwidth}
    \centering
    \includegraphics[scale=0.327]{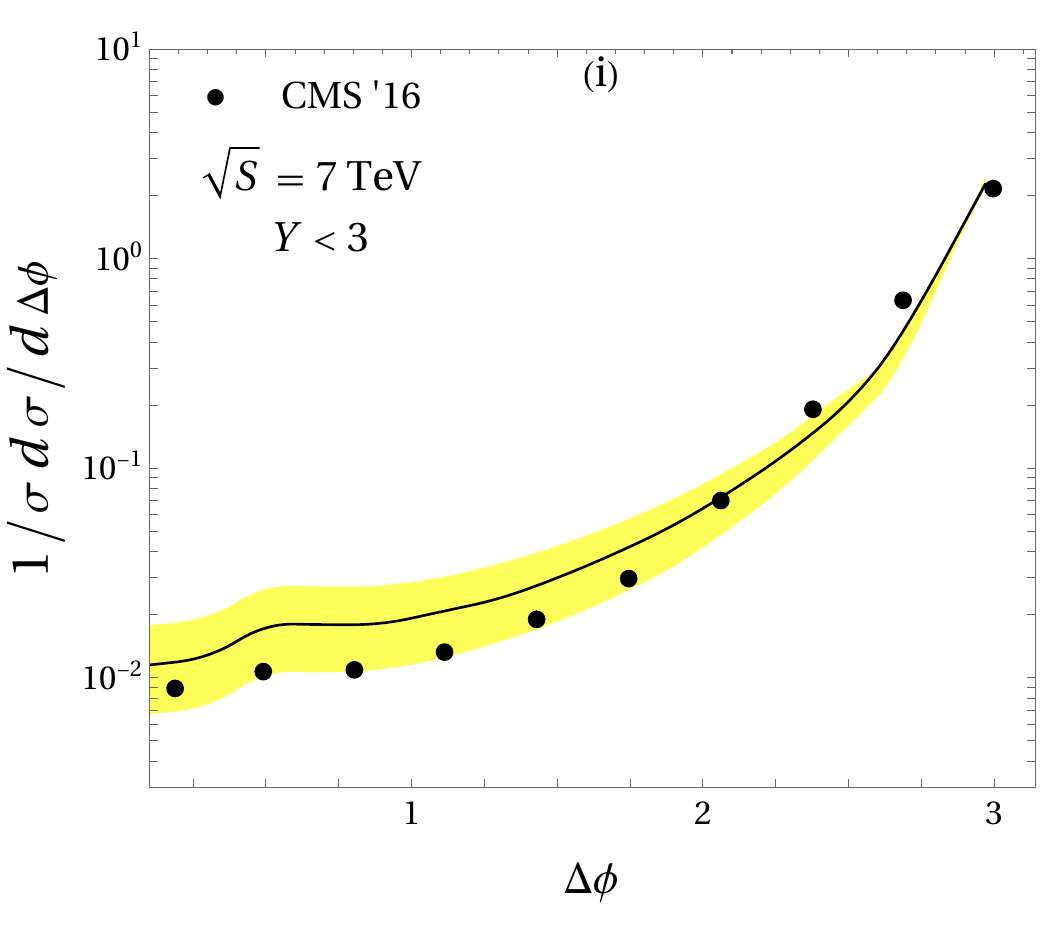}
    \includegraphics[scale=0.327]{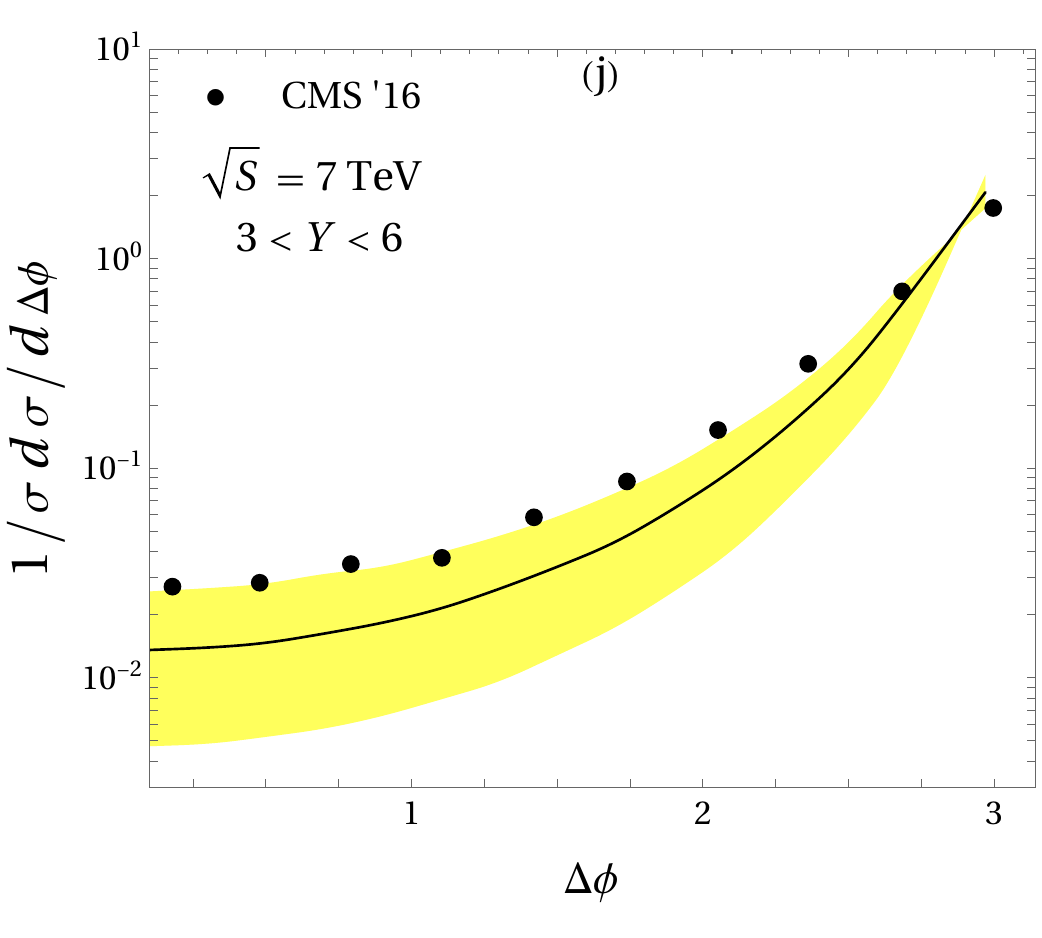}
    \includegraphics[scale=0.327]{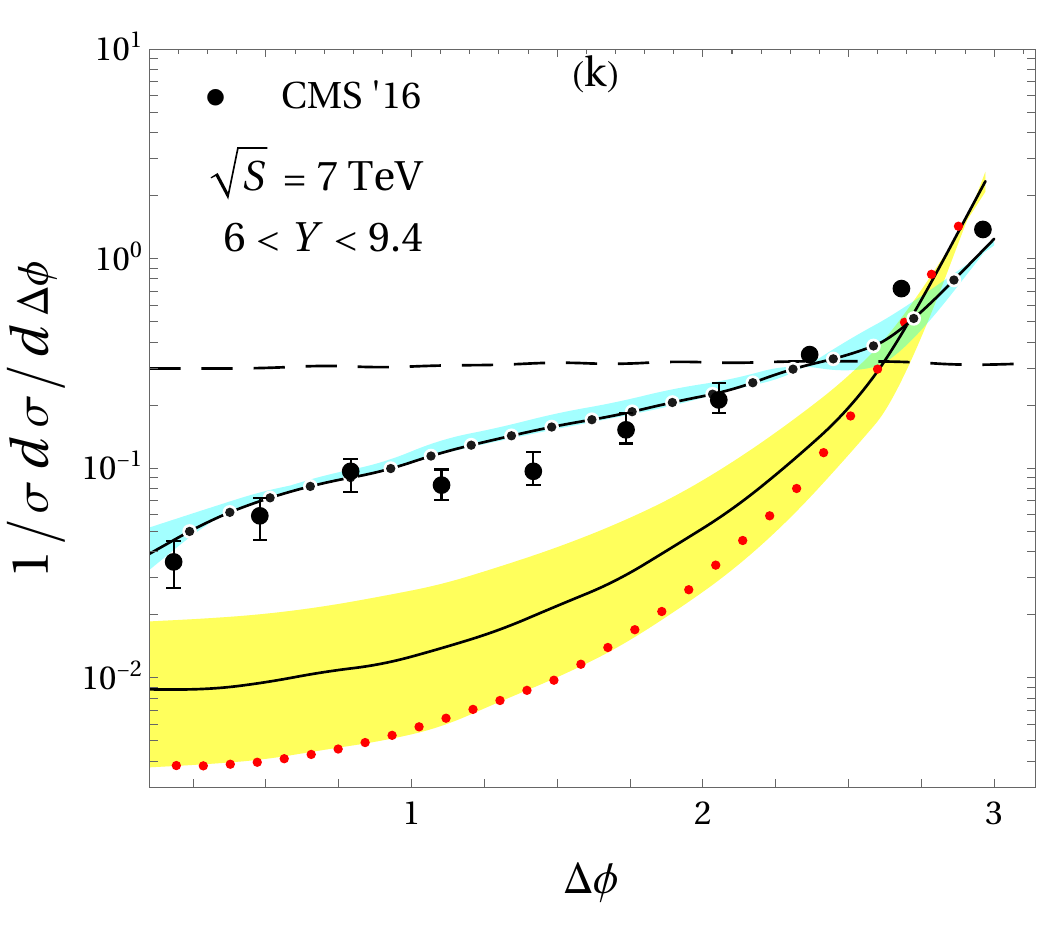}
\end{subfigure}
\\
\begin{subfigure}{0.32\textwidth}
    \raggedleft
    \includegraphics[scale=0.33]{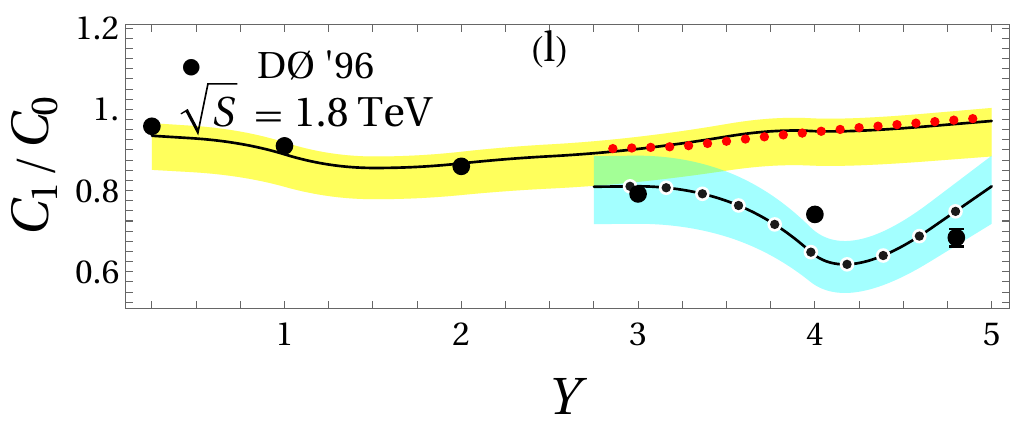}
    \vspace{0.15cm}
    \includegraphics[scale=0.33]{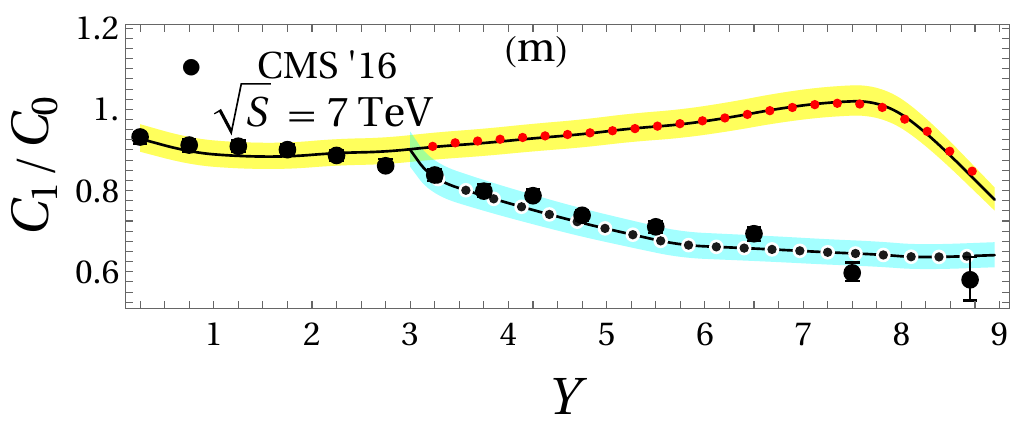}
\end{subfigure}
\begin{subfigure}{0.32\textwidth}
    \raggedleft
    \includegraphics[scale=0.33]{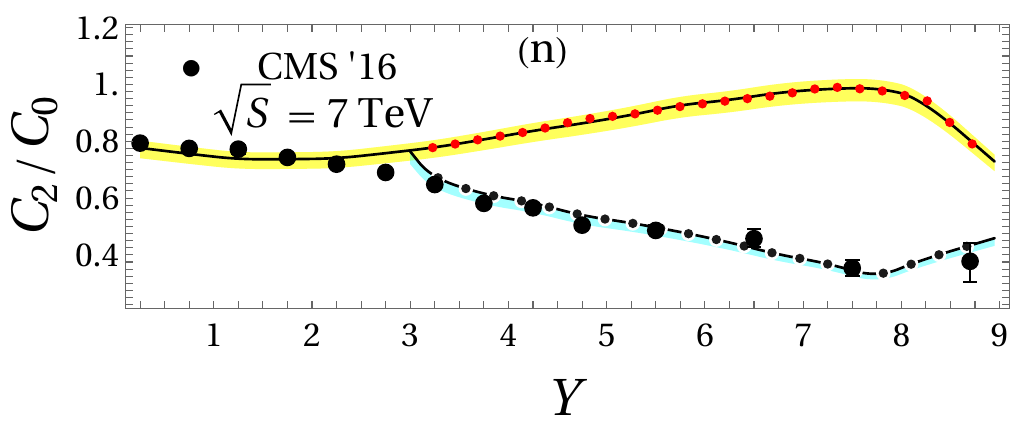}
    \vspace{0.15cm}
    \includegraphics[scale=0.33]{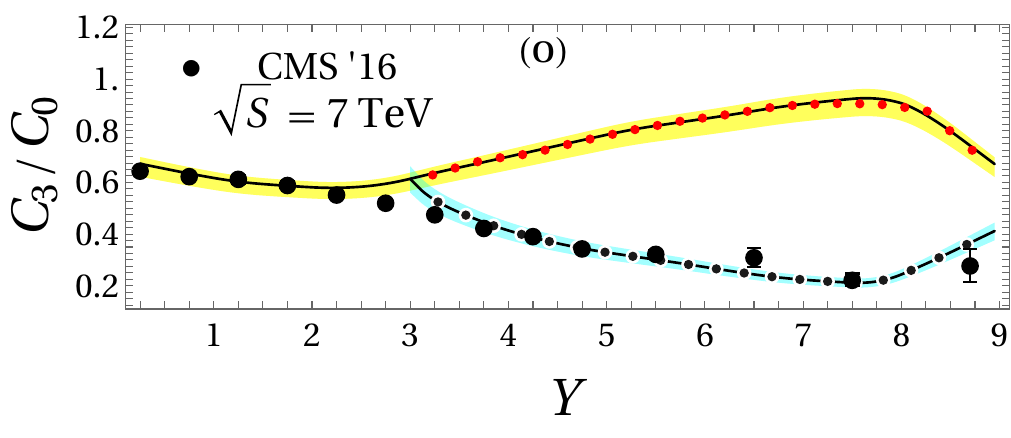}
\end{subfigure}
\begin{subfigure}{0.32\textwidth}
    \raggedleft
    \includegraphics[scale=0.33]{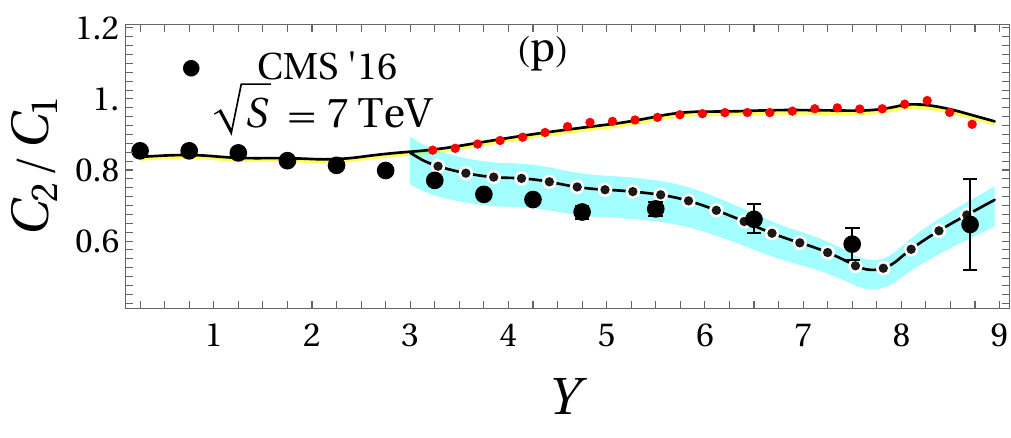}
    \vspace{0.15cm}
    \includegraphics[scale=0.33]{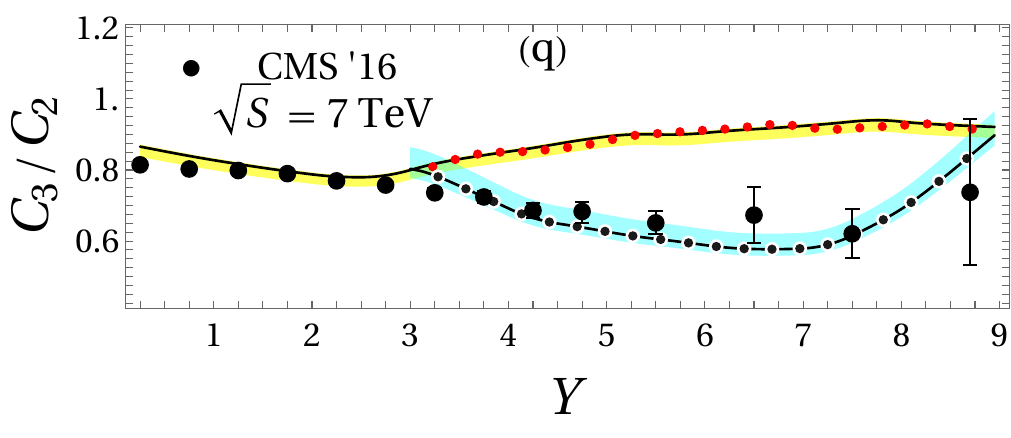}
\end{subfigure}
\caption{\raggedright Different observables related to the MN dijet production.
Several contributions are shown: HEF (solid line with yellow band), BFKL${}^{(0)}$ (red dotted line), LO BFKL (dashed line), and NLO BFKL (dot-dashed line with blue band).
The data are from CMS~\cite{CMS:1,CMS:2,CMS:3} and D$\O$~\cite{D0:1} Collaborations.}\label{fig:1}
\end{figure*}

Now we are in a position to compare the results of our computations with a subset of available data on the MN dijet production at hadron colliders.
Our predictions for the distributions over the rapidity difference $Y$ of the MN dijets produced in $p p$ collisions at $\sqrt{S} = 2.76$ TeV with $|{\bf p}_{\perp 1, 2}| > 35$ GeV~\cite{CMS:1} are shown in Fig.~\ref{fig:1}~(a).
The HEF contribution describes the experimental data well at low $Y < 2 - 3$ and gets subtracted by the BFKL${}^{(0)}$ contribution starting from  $Y > 3 - 3.5$, where the transition to the BFKL regime happens.
In the region of large $Y > 3 - 3.5$, the NLO BFKL-improved HEF computation is in agreement with the data, unlike the fixed-order HEF or the LO BFKL-improved HEF, clearly showing the importance of NLL BFKL corrections.
The uncertainties of the NLO BFKL-improved HEF with the Green's function calculated by the technique of Refs.~\cite{KC:13,KC:14} are of the same order as for the fixed-order HEF calculations, which demonstrates the  reduction of scale sensitivity of the calculations in this approach.
To make an unambiguous conclusion about manifestations of the BFKL effects, it is necessary to have data at higher LHC energies $\sqrt{S} = 7 - 13$ TeV and to consider the ratios $R$ of the cross sections at different energies, as it shown in Fig.~\ref{fig:1}~(b),~(c); at large $Y$, the NLO BFKL-improved HEF contribution grows significantly faster than the HEF one.

Another interesting observable is the ratio $R^{\rm MN}_{\rm ex}$ of the MN dijet cross section to a certain ``exclusive'' dijet cross section with the veto on extra jets, designed to suppress the ISR effects~\cite{CMS:1,CMS:3}.
In the HEF approach with the KMRW UPDFs, this ISR veto can be implemented by setting the upper limit on ${\bf l}_{\perp 1, 2}$ in Eq.~(\ref{eq:HEF}) as it was done in Ref.~\cite{PRA1}.
Our predictions at two energies, $\sqrt{S} = 2.76$ and $7$ TeV, compared to the experimental data are shown in Fig.~\ref{fig:1}~(d) and~(e) respectively.
The fixed-order HEF predictions are in satisfactory agreement with the data at small $Y < 3 - 3.5$, while for $Y > 3.5$ only the NLO BFKL-improved HEF predictions reproduce the data, whereas the LO BFKL-improved prediction gives a significantly lower value for the ratio.
This again demonstrates the importance of taking into account the NLL BFKL corrections.

It had been understood for a long time in collider physics that the observable most sensitive to the QCD radiation pattern, including the BFKL effects, is the correlation of the jets in azimuthal angle and corresponding ratios of angular harmonics~\cite{Stirling:94,SabioVera:07,SabioVera:11,Szymanowski:14,Papa:22}.
The NLO BFKL computation in the CF was found to be unstable at $\Delta\phi \to 0$ due to large $\ln {\bf l}_\perp'^2 / \mu_R^2$ in the NLO CF IFs.
To cure this instability, the application of the Sudakov resummation through the usage of the TMD factorization has been proposed~\cite{Szymanowski:16,Altinoluk:2023hfz}.
In our NLO BFKL-improved HEF computation we do the same resummation, since the improved KMRW uPDFs are consistent with the Sudakov resummation in the TMD factorization up to NLL level as shown in Ref.~\cite{PRA3}.
Moreover, due to the usage of the NLO eigenfunctions of Refs.~\cite{KC:13,KC:14} in our computations, we {\it do not} need to use the BLM scale-setting prescription, which was crucial for obtaining the agreement with the data in Refs.~\cite{Szymanowski:14,Papa:15}.

Our predictions for the azimuthal correlation spectra at $\sqrt{S} = 1.8$ and $7$ TeV and in different rapidity regions are shown in Fig.~\ref{fig:1} (f)--(k).
The kinematic cuts of the measurement by D$\O$~\cite{D0:1} are asymmetric, with minimum $|{\bf p}_\perp|$ of the leading and subleading jet respectively being $50$ GeV and $20$ GeV, while the cuts for CMS data~\cite{CMS:2} are symmetric: $|{\bf p}_{\perp 1, 2}| > 35$ GeV.
 At relatively small rapidities, the HEF contribution is enough to describe the data, see Fig.~\ref{fig:1}~(f),~(g),~(i),~(j).
The correct shape of the correlation spectra is evidence of the Sudakov resummation implemented in the KMRW UPDFs~(\ref{eq:C}).
At large $Y$, the HEF predicts stronger correlation between the MN jets than it is observed in the data, see Fig.~\ref{fig:1}~(h),~(k).
The necessary additional decorrelation comes from the radiations between the jets, which is taken into account by the BFKL resummation.
The LO BFKL-improved HEF predicts even too strong decorrelation at both $\sqrt{S} = 1.8$ and $7$ TeV, leading to an almost flat angular distribution.
A good description of the data can be achieved only by using the NLL BFKL resummation, which predicts the correct shape of the MN jets correlation spectra at large rapidities (Fig.~\ref{fig:1}~(h),~(k)).

The analysis of the angular distributions most sensitive to the BFKL dynamics is provided by the coefficients $C_k$ of the decomposition~\cite{Stirling:94,SabioVera:07,SabioVera:11,Szymanowski:14,Papa:22}:
$$
\frac{1}{\sigma} \frac{d \sigma}{d Y d \Delta\phi} =
\frac{1}{2 \pi} \left[ 1 + 2 \, \sum_{k \geq 0} \, C_k(Y) \cos\left( k ( \Delta\phi - \pi ) \right) \right],
$$
because the ratios of $C_k$ feature reduced scale-sensitivity in comparison with the absolute cross sections, as first observed in Ref.~\cite{SabioVera:11}.
Our predictions for the ratios of $C_k$ are presented in Fig.~\ref{fig:1}~(l)-(q), following our previous strategy.
The small-$Y$ region is covered by the HEF, which results in no suppression of the ratios $C_{k' > k} / C_k$ with the increasing of $Y$, in contrast to the Regge limit $Y \gg 1 / \als$, where the BFKL effects are dominant and Eq.~(\ref{eq:Grhs3}) predicts $C_{k' > k} / C_k \to 0$ with increasing $Y$.
This predictions are nicely confirmed in Fig.~\ref{fig:1} (l)--(q), where the experimental data for large $Y$ lie well below the HEF-only curve and are in agreement with the NLL BFKL-improved HEF curve.
The separation between these two curves in Fig.~\ref{fig:1} (l)--(q) reaches well above five standard deviations, even taking into account the residual scale-uncertainty bands, thus providing clear evidence in favour of the BFKL dynamics.

\section{Conclusions} \label{sec:conc}

To summarize, for the first time we have combined the resummation of the ISR effects provided by the HEF with the NLL BFKL resummation to describe the MN dijet hadroproduction.
The HEF describes the data well at low $Y$, while at large $Y$ the NLL BFKL resummation is necessary for the agreement with the data.
The adopted matching scheme smoothly interpolates predictions between the two descriptions.
We have used the technique of Refs.~\cite{KC:13,KC:14} to calculate the NLL BFKL Green's function, which reduces the renormalization scale uncertainty due to correct implementation of the RG-invariance at the NLL.
The usage of the HEF with the KMRW UPDFs, implementing the Sudakov resummation, together with the NLO solution for the BFKL Green's function eliminates the need for tuning of the scale choice in the MN dijet production process.
Thus, evidence for the manifestations of the BFKL effects becomes unambiguous, opening prospects for their future precision studies at BNL EIC, CERN HL-LHC and FCC, etc.

\section*{Acknowledgments}

Authors are grateful to A.~Kotikov, A.~Onishchenko, and M.~Fucilla for valuable feedback and V.~Fadin for drawing our attention to Ref.~\cite{AG:13}.
The work of A.C. and V.S. is supported by the BASIS Foundation for the Advancement of Theoretical Physics and Mathematics, grant \#24-1-1-16-5.
The work of M.N. is supported by binational Science Foundation grants \#2012124 and \#2021789, and by the ISF grant \#910/23.

\bibliography{References}

@article{BFKL1,
    author = "Lipatov, L. N.",
    title = "{Reggeization of the Vector Meson and the Vacuum Singularity in Nonabelian Gauge Theories}",
    journal = "Sov. J. Nucl. Phys.",
    volume = "23",
    pages = "338--345",
    year = "1976"
}

@article{BFKL2,
    author = "Kuraev, E. A. and Lipatov, L. N. and Fadin, Victor S.",
    title = "{Multi - Reggeon Processes in the Yang-Mills Theory}",
    journal = "Sov. Phys. JETP",
    volume = "44",
    pages = "443--450",
    year = "1976"
}

@article{BFKL3,
    author = "Kuraev, E. A. and Lipatov, L. N. and Fadin, Victor S.",
    title = "{The Pomeranchuk Singularity in Nonabelian Gauge Theories}",
    journal = "Sov. Phys. JETP",
    volume = "45",
    pages = "199--204",
    year = "1977"
}

@article{BFKL4,
    author = "Balitsky, I. I. and Lipatov, L. N.",
    title = "{The Pomeranchuk Singularity in Quantum Chromodynamics}",
    journal = "Sov. J. Nucl. Phys.",
    volume = "28",
    pages = "822--829",
    year = "1978"
}

@article{BFKL5,
    author = "Fadin, Victor S. and Lipatov, L. N.",
    title = "{BFKL pomeron in the next-to-leading approximation}",
    eprint = "hep-ph/9802290",
    archivePrefix = "arXiv",
    reportNumber = "DESY-98-033",
    doi = "10.1016/S0370-2693(98)00473-0",
    journal = "Phys. Lett. B",
    volume = "429",
    pages = "127--134",
    year = "1998"
}

@article{BFKL6,
    author = "Fadin, V. S. and Lipatov, L. N.",
    title = "{Reggeon cuts in QCD amplitudes with negative signature}",
    eprint = "1712.09805",
    archivePrefix = "arXiv",
    primaryClass = "hep-ph",
    reportNumber = "BUDKER-INP-2017-15",
    doi = "10.1140/epjc/s10052-018-5910-1",
    journal = "Eur. Phys. J. C",
    volume = "78",
    number = "6",
    pages = "439",
    year = "2018"
}

@article{EFT1,
    author = "Lipatov, L. N.",
    title = "{Gauge invariant effective action for high--energy processes in QCD}",
    reportNumber = "DESY-95-029",
    doi = "10.1016/0550-3213(95)00390-E",
    journal = "Nucl. Phys. B",
    volume = "452",
    pages = "369--400",
    year = "1995"
}

@article{EFT2,
    author = "Lipatov, L. N. and Vyazovsky, M. I.",
    title = "{QuasimultiRegge processes with a quark exchange in the $t$ channel}",
    doi = "10.1016/S0550-3213(00)00709-4",
    journal = "Nucl. Phys. B",
    volume = "597",
    pages = "399--409",
    year = "2001"
}

@article{EFT3,
    author = "Antonov, E. N. and Lipatov, L. N. and Kuraev, E. A. and Cherednikov, I. O.",
    title = "{Feynman rules for effective Regge action}",
    doi = "10.1016/j.nuclphysb.2005.013",
    journal = "Nucl. Phys. B",
    volume = "721",
    pages = "111--135",
    year = "2005"
}

@article{HEF1,
    author = "Gribov, L. V. and Levin, E. M. and Ryskin, M. G.",
    title = "{Semihard Processes in QCD}",
    doi = "10.1016/0370-1573(83)90022-4",
    journal = "Phys. Rept.",
    volume = "100",
    pages = "1--150",
    year = "1983"
}

@article{HEF2,
    author = "Catani, S. and Ciafaloni, M. and Hautmann, F.",
    title = "{GLUON CONTRIBUTIONS TO SMALL x HEAVY FLAVOR PRODUCTION}",
    reportNumber = "CAVENDISH-HEP-90-3",
    doi = "10.1016/0370-2693(90)91601-7",
    journal = "Phys. Lett. B",
    volume = "242",
    pages = "97--102",
    year = "1990"
}

@article{HEF3,
    author = "Catani, S. and Ciafaloni, M. and Hautmann, F.",
    title = "{High-energy factorization and small x heavy flavor production}",
    reportNumber = "CAVENDISH-HEP-90-27",
    doi = "10.1016/0550-3213(91)90055-3",
    journal = "Nucl. Phys. B",
    volume = "366",
    pages = "135--188",
    year = "1991"
}

@article{HEF4,
    author = "Catani, S. and Ciafaloni, M. and Hautmann, F.",
    title = "{High-energy factorization in QCD and minimal subtraction scheme}",
    reportNumber = "CERN-TH-6818-93",
    doi = "10.1016/0370-2693(93)90204-U",
    journal = "Phys. Lett. B",
    volume = "307",
    pages = "147--153",
    year = "1993"
}

@article{HEF5,
    author = "Catani, S. and Hautmann, F.",
    title = "{High-energy factorization and small x deep inelastic scattering beyond leading order}",
    eprint = "hep-ph/9405388",
    archivePrefix = "arXiv",
    reportNumber = "CAVENDISH-HEP-94-01",
    doi = "10.1016/0550-3213(94)90636-X",
    journal = "Nucl. Phys. B",
    volume = "427",
    pages = "475--524",
    year = "1994"
}

@article{HEF6,
    author = "Collins, John C. and Ellis, R. Keith",
    title = "{Heavy quark production in very high-energy hadron collisions}",
    reportNumber = "FERMILAB-PUB-91-022-T",
    doi = "10.1016/0550-3213(91)90288-9",
    journal = "Nucl. Phys. B",
    volume = "360",
    pages = "3--30",
    year = "1991"
}

@article{HN:25,
    author = "van Hameren, Andreas and Nefedov, Maxim",
    title = "{Hybrid high-energy factorization and evolution at NLO from the high-energy limit of collinear factorization}",
    eprint = "2501.02619",
    archivePrefix = "arXiv",
    primaryClass = "hep-ph",
    reportNumber = "IFJPAN-IV-2025-1",
    doi = "10.1007/JHEP02(2025)160",
    journal = "JHEP",
    volume = "02",
    pages = "160",
    year = "2025"
}

@article{PRA1,
    author = "Nefedov, M.A. and Saleev, V.A. and Shipilova, A.V",
    title = "{Dijet azimuthal decorrelations at the LHC in the parton Reggeization approach}",
    reportNumber = "DESY-13-070, PREPRINT-DESY-13-070",
    doi = "10.1103/PhysRevD.87.094030",
    journal = "Phys. Rev. D",
    volume = "87",
    number = "9",
    pages = "094030",
    year = "2013"
}

@article{PRA3,
    author = "Nefedov, M.A. and Saleev, V.A.",
    title = "{High-Energy Factorization for Drell-Yan process in $pp$ and $p{\bar p}$ collisions with new Unintegrated PDFs}",
    doi = "10.1103/PhysRevD.102.114018",
    journal = "Phys. Rev. D",
    volume = "102",
    pages = "114018",
    year = "2020"
}

@article{MN:21,
    author = "Nefedov, Maxim",
    title = "{Sudakov resummation from the BFKL evolution}",
    eprint = "2105.13915",
    archivePrefix = "arXiv",
    primaryClass = "hep-ph",
    doi = "10.1103/PhysRevD.104.054039",
    journal = "Phys. Rev. D",
    volume = "104",
    number = "5",
    pages = "054039",
    year = "2021"
}

@article{HKNS:19,
    author = "He, Zhi-Guo and Kniehl, B.A. and Nefedov, M.A. and Saleev, V.A.",
    title = "{Double Prompt $J/\psi$ Hadroproduction in the Parton Reggeization Approach with High-Energy Resummation}",
    eprint = "1906.08979",
    archivePrefix = "arXiv",
    primaryClass = "hep-ph",
    reportNumber = "DESY 19-111, DESY-19-111",
    doi = "10.1103/PhysRevLett.123.162002",
    journal = "Phys. Rev. Lett.",
    volume = "123",
    number = "16",
    pages = "162002",
    year = "2019"
}

@book{KL:12,
    author = "Kovchegov, Y.V. and Levin, E.",
    title = "{Quantum Chromodynamics at High Energy}",
    doi = "10.1017/9781009291446",
    isbn = "978-1-009-29144-6, 978-1-009-29141-5, 978-1-009-29142-2, 978-0-521-11257-4, 978-1-139-55768-9",
    publisher = "Oxford University Press",
    volume = "33",
    year = "2013"
}

@book{LFI:10,
    author = "Ioffe, B.L. and Fadin, V.S. and Lipatov, L.N.",
    title = "{Quantum chromodynamics: Perturbative and nonperturbative aspects}",
    doi = "10.1017/CBO9780511711817",
    isbn = "978-1-107-42475-3, 978-0-521-63148-8, 978-0-511-71744-4",
    publisher = "Cambridge Univ. Press",
    year = "2010"
}

@article{KL:00,
    author = "Kotikov, A. V. and Lipatov, L. N.",
    title = "{NLO corrections to the BFKL equation in QCD and in supersymmetric gauge theories}",
    eprint = "hep-ph/0004008",
    archivePrefix = "arXiv",
    reportNumber = "DESY-00-059",
    doi = "10.1016/S0550-3213(00)00329-1",
    journal = "Nucl. Phys. B",
    volume = "582",
    pages = "19--43",
    year = "2000"
}

@article{KC:13,
    author = "Chirilli, Giovanni A. and Kovchegov, Yuri V.",
    title = "{Solution of the NLO BFKL Equation and a Strategy for Solving the All-Order BFKL Equation}",
    eprint = "1305.1924",
    archivePrefix = "arXiv",
    primaryClass = "hep-ph",
    doi = "10.1007/JHEP06(2013)055",
    journal = "JHEP",
    volume = "06",
    pages = "055",
    year = "2013"
}

@article{KC:14,
    author = "Chirilli, Giovanni A. and Kovchegov, Yuri V.",
    title = "{$\gamma^* \gamma^*$ Cross Section at NLO and Properties of the BFKL Evolution at Higher Orders}",
    eprint = "1403.3384",
    archivePrefix = "arXiv",
    primaryClass = "hep-ph",
    doi = "10.1007/JHEP05(2014)099",
    journal = "JHEP",
    volume = "05",
    pages = "099",
    year = "2014",
    note = "[Erratum: JHEP 08, 075 (2015)]"
}

@article{MN:87,
    author = "Mueller, Alfred H. and Navelet, H.",
    title = "{An Inclusive Minijet Cross-Section and the Bare Pomeron in QCD}",
    reportNumber = "SACLAY-SPHT-86-094",
    doi = "10.1016/0550-3213(87)90705-X",
    journal = "Nucl. Phys. B",
    volume = "282",
    pages = "727--744",
    year = "1987"
}

@article{KMR,
    author = "Kimber, M. A. and Martin, Alan D. and Ryskin, M. G.",
    title = "{Unintegrated parton distributions}",
    doi = "10.1103/PhysRevD.63.114027",
    journal = "Phys. Rev. D",
    volume = "63",
    pages = "114027",
    year = "2001"
}

@article{MRW,
    author = "Watt, G. and Martin, A. D. and Ryskin, M. G.",
    title = "{Unintegrated parton distributions and inclusive jet production at HERA}",
    doi = "10.1140/epjc/s2003-01320-4",
    journal = "Eur. Phys. J. C",
    volume = "31",
    pages = "73--89",
    year = "2003"
}

@article{CUBA,
    author = "Hahn, T.",
    title = "{CUBA: A Library for multidimensional numerical integration}",
    eprint = "hep-ph/0404043",
    archivePrefix = "arXiv",
    reportNumber = "MPP-2004-40",
    doi = "10.1016/j.cpc.2005.01.010",
    journal = "Comput. Phys. Commun.",
    volume = "168",
    pages = "78--95",
    year = "2005"
}

@article{MSTW,
    author = "Martin, A. D. and Stirling, W. J. and Thorne, R. S. and Watt, G.",
    title = "{Parton distributions for the LHC}",
    eprint = "0901.0002",
    archivePrefix = "arXiv",
    primaryClass = "hep-ph",
    reportNumber = "IPPP-08-95, DCPT-08-190, CAVENDISH-HEP-08-16",
    doi = "10.1140/epjc/s10052-009-1072-5",
    journal = "Eur. Phys. J. C",
    volume = "63",
    pages = "189--285",
    year = "2009"
}

@article{Hentschinski:22,
    author = "Hentschinski, Martin and others",
    title = "{White Paper on Forward Physics, BFKL, Saturation Physics and Diffraction}",
    eprint = "2203.08129",
    archivePrefix = "arXiv",
    primaryClass = "hep-ph",
    doi = "10.5506/APhysPolB.54.3-A2",
    journal = "Acta Phys. Polon. B",
    volume = "54",
    number = "3",
    pages = "3-A2",
    year = "2023"
}

@article{DelDuca:93,
    author = "Del Duca, Vittorio and Schmidt, Carl R.",
    title = "{Dijet production at large rapidity intervals}",
    eprint = "hep-ph/9311290",
    archivePrefix = "arXiv",
    reportNumber = "DESY-93-139, SCIPP-93-35",
    doi = "10.1103/PhysRevD.49.4510",
    journal = "Phys. Rev. D",
    volume = "49",
    pages = "4510--4516",
    year = "1994"
}

@article{Stirling:94,
    author = "Stirling, W. James",
    title = "{Production of jet pairs at large relative rapidity in hadron hadron collisions as a probe of the perturbative pomeron}",
    eprint = "hep-ph/9401266",
    archivePrefix = "arXiv",
    reportNumber = "DTP-94-04",
    doi = "10.1016/0550-3213(94)90565-7",
    journal = "Nucl. Phys. B",
    volume = "423",
    pages = "56--79",
    year = "1994"
}

@article{Andersen:01,
    author = "Andersen, J. R. and Del Duca, V. and Frixione, S. and Schmidt, C. R. and Stirling, W. James",
    title = "{Mueller-Navelet jets at hadron colliders}",
    eprint = "hep-ph/0101180",
    archivePrefix = "arXiv",
    reportNumber = "IPPP-00-04, DTP-00-64, MSUHEP-01024, DFTT-44-200",
    doi = "10.1088/1126-6708/2001/02/007",
    journal = "JHEP",
    volume = "02",
    pages = "007",
    year = "2001"
}

@article{SabioVera:05,
    author = "Sabio Vera, Agustin",
    title = "{An 'All-poles' approximation to collinear resummations in the Regge limit of perturbative QCD}",
    eprint = "hep-ph/0505128",
    archivePrefix = "arXiv",
    reportNumber = "DESY-05-088",
    doi = "10.1016/j.nuclphysb.2005.06.003",
    journal = "Nucl. Phys. B",
    volume = "722",
    pages = "65--80",
    year = "2005"
}

@article{SabioVera:07,
    author = "Sabio Vera, Agustin and Schwennsen, Florian",
    title = "{The Azimuthal decorrelation of jets widely separated in rapidity as a test of the BFKL kernel}",
    eprint = "hep-ph/0702158",
    archivePrefix = "arXiv",
    reportNumber = "CERN-PH-TH-2007-029, DESY-07-017",
    doi = "10.1016/j.nuclphysb.2007.03.050",
    journal = "Nucl. Phys. B",
    volume = "776",
    pages = "170--186",
    year = "2007"
}

@article{SabioVera:11,
    author = "Angioni, M. and Chachamis, G. and Madrigal, J. D. and Sabio Vera, A.",
    title = "{Dijet Production at Large Rapidity Separation in N=4 SYM}",
    eprint = "1106.6172",
    archivePrefix = "arXiv",
    primaryClass = "hep-th",
    reportNumber = "LPN11-30, IFT-UAM-CSIC-11-38, FTUAM-11-48",
    doi = "10.1103/PhysRevLett.107.191601",
    journal = "Phys. Rev. Lett.",
    volume = "107",
    pages = "191601",
    year = "2011"
}

@article{Szymanowski:10,
    author = "Colferai, D. and Schwennsen, F. and Szymanowski, L. and Wallon, S.",
    title = "{Mueller Navelet jets at LHC - complete NLL BFKL calculation}",
    eprint = "1002.1365",
    archivePrefix = "arXiv",
    primaryClass = "hep-ph",
    reportNumber = "CPHT-RR139.1209, LPT-ORSAY-09-111, DFF-452-10-2009",
    doi = "10.1007/JHEP12(2010)026",
    journal = "JHEP",
    volume = "12",
    pages = "026",
    year = "2010"
}

@article{Szymanowski:13,
    author = "Ducloue, B. and Szymanowski, L. and Wallon, S.",
    title = "{Confronting Mueller-Navelet jets in NLL BFKL with LHC experiments at 7 TeV}",
    eprint = "1302.7012",
    archivePrefix = "arXiv",
    primaryClass = "hep-ph",
    reportNumber = "LPT-ORSAY-13-18",
    doi = "10.1007/JHEP05(2013)096",
    journal = "JHEP",
    volume = "05",
    pages = "096",
    year = "2013"
}

@article{Szymanowski:14,
    author = "Duclou\'e, B. and Szymanowski, L. and Wallon, S.",
    title = "{Evidence for high-energy resummation effects in Mueller-Navelet jets at the LHC}",
    eprint = "1309.3229",
    archivePrefix = "arXiv",
    primaryClass = "hep-ph",
    reportNumber = "LPT-ORSAY-13-71",
    doi = "10.1103/PhysRevLett.112.082003",
    journal = "Phys. Rev. Lett.",
    volume = "112",
    pages = "082003",
    year = "2014"
}

@article{Szymanowski:15,
    author = "Duclou\'e, B. and Szymanowski, L. and Wallon, S.",
    title = "{Evaluating the double parton scattering contribution to Mueller-Navelet jets production at the LHC}",
    eprint = "1507.04735",
    archivePrefix = "arXiv",
    primaryClass = "hep-ph",
    reportNumber = "LPT-ORSAY-15-52",
    doi = "10.1103/PhysRevD.92.076002",
    journal = "Phys. Rev. D",
    volume = "92",
    number = "7",
    pages = "076002",
    year = "2015"
}

@article{Szymanowski:16,
    author = "Mueller, A. H. and Szymanowski, Lech and Wallon, Samuel and Xiao, Bo-Wen and Yuan, Feng",
    title = "{Sudakov Resummations in Mueller-Navelet Dijet Production}",
    eprint = "1512.07127",
    archivePrefix = "arXiv",
    primaryClass = "hep-ph",
    reportNumber = "LPT-ORSAY-15-85",
    doi = "10.1007/JHEP03(2016)096",
    journal = "JHEP",
    volume = "03",
    pages = "096",
    year = "2016"
}

@article{Papa:13,
    author = "Caporale, Francesco and Ivanov, Dmitry Yu. and Murdaca, Beatrice and Papa, Alessandro",
    title = "{Mueller-Navelet small-cone jets at LHC in next-to-leading BFKL}",
    eprint = "1211.7225",
    archivePrefix = "arXiv",
    primaryClass = "hep-ph",
    doi = "10.1016/j.nuclphysb.2013.09.013",
    journal = "Nucl. Phys. B",
    volume = "877",
    pages = "73--94",
    year = "2013"
}

@article{Papa:14,
    author = "Caporale, Francesco and Ivanov, Dmitry Yu. and Murdaca, Beatrice and Papa, Alessandro",
    title = "{Mueller\textendash{}Navelet jets in next-to-leading order BFKL: theory versus experiment}",
    eprint = "1407.8431",
    archivePrefix = "arXiv",
    primaryClass = "hep-ph",
    doi = "10.1140/epjc/s10052-015-3754-5",
    journal = "Eur. Phys. J. C",
    volume = "74",
    number = "10",
    pages = "3084",
    year = "2014",
    note = "[Erratum: Eur.Phys.J.C 75, 535 (2015)]"
}

@article{Papa:15,
    author = "Caporale, Francesco and Ivanov, Dmitry Yu. and Murdaca, Beatrice and Papa, Alessandro",
    title = "{Brodsky-Lepage-Mackenzie optimal renormalization scale setting for semihard processes}",
    eprint = "1504.06471",
    archivePrefix = "arXiv",
    primaryClass = "hep-ph",
    doi = "10.1103/PhysRevD.91.114009",
    journal = "Phys. Rev. D",
    volume = "91",
    number = "11",
    pages = "114009",
    year = "2015"
}

@article{Papa:22,
    author = "Celiberto, Francesco Giovanni and Papa, Alessandro",
    title = "{Mueller-Navelet jets at the LHC: Hunting data with azimuthal distributions}",
    eprint = "2207.05015",
    archivePrefix = "arXiv",
    primaryClass = "hep-ph",
    doi = "10.1103/PhysRevD.106.114004",
    journal = "Phys. Rev. D",
    volume = "106",
    number = "11",
    pages = "114004",
    year = "2022"
}

@article{Kim:23,
    author = "Egorov, Anatolii Iu. and Kim, Victor T.",
    title = "{Next-to-leading BFKL evolution for dijets with large rapidity separation at different LHC energies}",
    eprint = "2305.19854",
    archivePrefix = "arXiv",
    primaryClass = "hep-ph",
    doi = "10.1103/PhysRevD.108.014010",
    journal = "Phys. Rev. D",
    volume = "108",
    number = "1",
    pages = "014010",
    year = "2023"
}

@article{D0:1,
    author = "Abachi, S. and others",
    collaboration = "D0",
    title = "{The Azimuthal decorrelation of jets widely separated in rapidity}",
    eprint = "hep-ex/9603010",
    archivePrefix = "arXiv",
    reportNumber = "FERMILAB-PUB-96-038-E",
    doi = "10.1103/PhysRevLett.77.595",
    journal = "Phys. Rev. Lett.",
    volume = "77",
    pages = "595--600",
    year = "1996"
}

@article{CMS:1,
    author = "Chatrchyan, Serguei and others",
    collaboration = "CMS",
    title = "{Ratios of Dijet Production Cross Sections as a Function of the Absolute Difference in Rapidity between Jets in Proton-Proton Collisions at $\sqrt{s}=7$ TeV}",
    eprint = "1204.0696",
    archivePrefix = "arXiv",
    primaryClass = "hep-ex",
    reportNumber = "CMS-FWD-10-014, CERN-PH-EP-2012-088",
    doi = "10.1140/epjc/s10052-012-2216-6",
    journal = "Eur. Phys. J. C",
    volume = "72",
    pages = "2216",
    year = "2012"
}

@article{CMS:2,
    author = "Khachatryan, Vardan and others",
    collaboration = "CMS",
    title = "{Azimuthal decorrelation of jets widely separated in rapidity in pp collisions at $ \sqrt{s}=7 $ TeV}",
    eprint = "1601.06713",
    archivePrefix = "arXiv",
    primaryClass = "hep-ex",
    reportNumber = "CMS-FSQ-12-002, CERN-PH-EP-2015-309",
    doi = "10.1007/JHEP08(2016)139",
    journal = "JHEP",
    volume = "08",
    pages = "139",
    year = "2016"
}

@article{CMS:3,
    author = "Tumasyan, Armen and others",
    collaboration = "CMS",
    title = "{Study of dijet events with large rapidity separation in proton-proton collisions at $ \sqrt{s} $ = 2.76 TeV}",
    eprint = "2111.04605",
    archivePrefix = "arXiv",
    primaryClass = "hep-ex",
    reportNumber = "CMS-FSQ-13-004, CERN-EP-2021-173",
    doi = "10.1007/JHEP03(2022)189",
    journal = "JHEP",
    volume = "03",
    pages = "189",
    year = "2022"
}

@article{Salam:98,
    author = "Salam, G. P.",
    title = "{A Resummation of large subleading corrections at small x}",
    eprint = "hep-ph/9806482",
    archivePrefix = "arXiv",
    reportNumber = "IFUM-627-FT",
    doi = "10.1088/1126-6708/1998/07/019",
    journal = "JHEP",
    volume = "07",
    pages = "019",
    year = "1998"
}

@article{BLM,
    author = "Brodsky, Stanley J. and Lepage, G. Peter and Mackenzie, Paul B.",
    title = "{On the Elimination of Scale Ambiguities in Perturbative Quantum Chromodynamics}",
    reportNumber = "SLAC-PUB-3011, FERMILAB-PUB-83-040-T",
    doi = "10.1103/PhysRevD.28.228",
    journal = "Phys. Rev. D",
    volume = "28",
    pages = "228",
    year = "1983"
}

@article{BFKLP,
    author = "Brodsky, Stanley J. and Fadin, Victor S. and Kim, Victor T. and Lipatov, Lev N. and Pivovarov, Grigorii B.",
    title = "{The QCD pomeron with optimal renormalization}",
    eprint = "hep-ph/9901229",
    archivePrefix = "arXiv",
    reportNumber = "SLAC-PUB-8037, IITAP-98-010",
    doi = "10.1134/1.568145",
    journal = "JETP Lett.",
    volume = "70",
    pages = "155--160",
    year = "1999"
}

@article{Fadin:2023roz,
    author = "Fadin, Victor S. and Fucilla, Michael and Papa, Alessandro",
    title = "{One-loop Lipatov vertex in QCD with higher \ensuremath{\epsilon}-accuracy}",
    eprint = "2302.09868",
    archivePrefix = "arXiv",
    primaryClass = "hep-ph",
    doi = "10.1007/JHEP04(2023)137",
    journal = "JHEP",
    volume = "04",
    pages = "137",
    year = "2023"
}

@article{Byrne:2022wzk,
    author = "Byrne, Emmet P. and Del Duca, Vittorio and Dixon, Lance J. and Gardi, Einan and Smillie, Jennifer M.",
    title = "{One-loop central-emission vertex for two gluons in $ \mathcal{N} $ = 4 super Yang-Mills theory}",
    eprint = "2204.12459",
    archivePrefix = "arXiv",
    primaryClass = "hep-ph",
    reportNumber = "SLAC-PUB-17654",
    doi = "10.1007/JHEP08(2022)271",
    journal = "JHEP",
    volume = "08",
    pages = "271",
    year = "2022"
}

@article{Caola:2021izf,
    author = "Caola, Fabrizio and Chakraborty, Amlan and Gambuti, Giulio and von Manteuffel, Andreas and Tancredi, Lorenzo",
    title = "{Three-Loop Gluon Scattering in QCD and the Gluon Regge Trajectory}",
    eprint = "2112.11097",
    archivePrefix = "arXiv",
    primaryClass = "hep-ph",
    reportNumber = "OUTP-21-28P, MSUHEP-21-035, TUM-HEP-1382/21",
    doi = "10.1103/PhysRevLett.128.212001",
    journal = "Phys. Rev. Lett.",
    volume = "128",
    number = "21",
    pages = "212001",
    year = "2022"
}

@article{Falcioni:2021dgr,
    author = "Falcioni, Giulio and Gardi, Einan and Maher, Niamh and Milloy, Calum and Vernazza, Leonardo",
    title = "{Disentangling the Regge Cut and Regge Pole in Perturbative QCD}",
    eprint = "2112.11098",
    archivePrefix = "arXiv",
    primaryClass = "hep-ph",
    reportNumber = "CERN-TH-2021-225",
    doi = "10.1103/PhysRevLett.128.132001",
    journal = "Phys. Rev. Lett.",
    volume = "128",
    number = "13",
    pages = "132001",
    year = "2022"
}

@article{DelDuca:2021vjq,
    author = "Del Duca, Vittorio and Marzucca, Robin and Verbeek, Bram",
    title = "{The gluon Regge trajectory at three loops from planar Yang-Mills theory}",
    eprint = "2111.14265",
    archivePrefix = "arXiv",
    primaryClass = "hep-ph",
    reportNumber = "UUIPT-59/21",
    doi = "10.1007/JHEP01(2022)149",
    journal = "JHEP",
    volume = "01",
    pages = "149",
    year = "2022"
}

@article{Byrne:2023nqx,
    author = "Byrne, Emmet P.",
    title = "{One-loop five-parton amplitudes in the NMRK limit}",
    eprint = "2312.15051",
    archivePrefix = "arXiv",
    primaryClass = "hep-ph",
    doi = "10.1007/JHEP07(2024)284",
    journal = "JHEP",
    volume = "07",
    pages = "284",
    year = "2024"
}

@article{Abreu:2024xoh,
    author = "Abreu, Samuel and De Laurentis, Giuseppe and Falcioni, Giulio and Gardi, Einan and Milloy, Calum and Vernazza, Leonardo",
    title = "{The two-loop Lipatov vertex in QCD}",
    eprint = "2412.20578",
    archivePrefix = "arXiv",
    primaryClass = "hep-ph",
    reportNumber = "CERN-TH-2024-226, ZU-TH 68/24",
    doi = "10.1007/JHEP04(2025)161",
    journal = "JHEP",
    volume = "04",
    pages = "161",
    year = "2025"
}

@article{Polizzi:2025edm,
    author = "Polizzi, Ada and Fucilla, Michael and Papa, Alessandro",
    title = "{High-energy factorization via eigenfunctions of the next-to-leading-order BFKL kernel}",
    eprint = "2505.18074",
    archivePrefix = "arXiv",
    primaryClass = "hep-ph",
    month = "5",
    year = "2025"
}

@article{Buccioni:2024gzo,
    author = "Buccioni, Federico and Caola, Fabrizio and Devoto, Federica and Gambuti, Giulio",
    title = "{Investigating the universality of five-point QCD scattering amplitudes at high energy}",
    eprint = "2411.14050",
    archivePrefix = "arXiv",
    primaryClass = "hep-ph",
    reportNumber = "OUTP-24-06P, SLAC-PUB-241120, TUM-HEP-1537/24",
    doi = "10.1007/JHEP03(2025)129",
    journal = "JHEP",
    volume = "03",
    pages = "129",
    year = "2025"
}

@article{AG:13,
    author = "Grabovsky, A. V.",
    title = "{On the solution to the NLO forward BFKL equation}",
    eprint = "1307.3152",
    archivePrefix = "arXiv",
    primaryClass = "hep-ph",
    doi = "10.1007/JHEP09(2013)098",
    journal = "JHEP",
    volume = "09",
    pages = "098",
    year = "2013"
}

@article{AbdulKhalek:2021gbh,
    author = "Abdul Khalek, R. and others",
    title = "{Science Requirements and Detector Concepts for the Electron-Ion Collider}: {EIC Yellow Report}",
    eprint = "2103.05419",
    archivePrefix = "arXiv",
    primaryClass = "physics.ins-det",
    reportNumber = "BNL-220990-2021-FORE, JLAB-PHY-21-3198, LA-UR-21-20953",
    doi = "10.1016/j.nuclphysa.2022.122447",
    journal = "Nucl. Phys. A",
    volume = "1026",
    pages = "122447",
    year = "2022"
}

@article{Baldenegro:2024ndr,
    author = "Baldenegro, C. and Chachamis, G. and Kampshoff, M. and Klasen, M. and Milhano, G. J. and Royon, C. and Sabio Vera, A.",
    title = "{Multijet event shape variables for Mueller-Navelet jet topologies}",
    eprint = "2406.10681",
    archivePrefix = "arXiv",
    primaryClass = "hep-ph",
    doi = "10.1103/PhysRevD.110.114027",
    journal = "Phys. Rev. D",
    volume = "110",
    number = "11",
    pages = "114027",
    year = "2024"
}

@article{Altinoluk:2023hfz,
    author = "Altinoluk, Tolga and Armesto, N{\'e}stor and Kovner, Alexander and Lublinsky, Michael",
    title = "{Single inclusive particle production at next-to-leading order in proton-nucleus collisions at forward rapidities: Hybrid approach meets TMD factorization}",
    eprint = "2307.14922",
    archivePrefix = "arXiv",
    primaryClass = "hep-ph",
    doi = "10.1103/PhysRevD.108.074003",
    journal = "Phys. Rev. D",
    volume = "108",
    number = "7",
    pages = "074003",
    year = "2023"
}

@article{CSS,
    author = "Collins, John C. and Soper, Davison E. and Sterman, George F.",
    title = "{Transverse Momentum Distribution in Drell-Yan Pair and W and Z Boson Production}",
    reportNumber = "CERN-TH-3923",
    doi = "10.1016/0550-3213(85)90479-1",
    journal = "Nucl. Phys. B",
    volume = "250",
    pages = "199--224",
    year = "1985"
}

@inproceedings{Blumlein:1995eu,
      author         = "Blumlein, Johannes",
      title          = "{On the k(T) dependent gluon density of the proton}",
      booktitle      = "{Deep inelastic scattering and QCD. Proceedings,
                        Workshop, Paris, France, April 24-28, 1995}",
      url            = "http://www-library.desy.de/cgi-bin/showprep.pl?desy95-121",
      year           = "1995",
      pages          = "265-268",
      eprint         = "hep-ph/9506403",
      archivePrefix  = "arXiv",
      primaryClass   = "hep-ph",
      reportNumber   = "DESY-95-121",
      SLACcitation   = "%%CITATION = HEP-PH/9506403;%%"
}

@article{Ciafaloni:2003rd,
    author = "Ciafaloni, M. and Colferai, D. and Salam, G. P. and Stasto, A. M.",
    title = "{Renormalization group improved small x Green's function}",
    eprint = "hep-ph/0307188",
    archivePrefix = "arXiv",
    reportNumber = "DESY-03-060, DFF-404-05-03, LPTHE-03-20",
    doi = "10.1103/PhysRevD.68.114003",
    journal = "Phys. Rev. D",
    volume = "68",
    pages = "114003",
    year = "2003"
}

@article{DelDuca:2001gu,
    author = "Del Duca, Vittorio and Glover, E. W. Nigel",
    title = "{The High-energy limit of QCD at two loops}",
    eprint = "hep-ph/0109028",
    archivePrefix = "arXiv",
    reportNumber = "DCPT-01-66, IPPP-01-33, DFTT-25-2001",
    doi = "10.1088/1126-6708/2001/10/035",
    journal = "JHEP",
    volume = "10",
    pages = "035",
    year = "2001"
}

@article{Caron-Huot:2013fea,
    author = "Caron-Huot, Simon",
    title = "{When does the gluon reggeize?}",
    eprint = "1309.6521",
    archivePrefix = "arXiv",
    primaryClass = "hep-th",
    doi = "10.1007/JHEP05(2015)093",
    journal = "JHEP",
    volume = "05",
    pages = "093",
    year = "2015"
}

@article{Lansberg:2021vie,
    author = "Lansberg, Jean-Philippe and Nefedov, Maxim and Ozcelik, Melih A.",
    title = "{Matching next-to-leading-order and high-energy-resummed calculations of heavy-quarkonium-hadroproduction cross sections}",
    eprint = "2112.06789",
    archivePrefix = "arXiv",
    primaryClass = "hep-ph",
    reportNumber = "TTP-21-057",
    doi = "10.1007/JHEP05(2022)083",
    journal = "JHEP",
    volume = "05",
    pages = "083",
    year = "2022"
}

@article{Lansberg:2023kzf,
    author = "Lansberg, Jean-Philippe and Nefedov, Maxim and Ozcelik, Melih A.",
    title = "{Curing the high-energy perturbative instability of vector-quarkonium-photoproduction cross sections at order $\alpha \alpha _s^3$ with high-energy factorisation}",
    eprint = "2306.02425",
    archivePrefix = "arXiv",
    primaryClass = "hep-ph",
    doi = "10.1140/epjc/s10052-024-12588-x",
    journal = "Eur. Phys. J. C",
    volume = "84",
    number = "4",
    pages = "351",
    year = "2024"
}

@article{Flett:2024htj,
    author = "Flett, C. A. and Lansberg, J. P. and Nabeebaccus, S. and Nefedov, M. and Sznajder, P. and Wagner, J.",
    title = "{Exclusive vector-quarkonium photoproduction at NLO in {\ensuremath{\alpha}}s in collinear factorisation with evolution of the generalised parton distributions and high-energy resummation}",
    eprint = "2409.05738",
    archivePrefix = "arXiv",
    primaryClass = "hep-ph",
    doi = "10.1016/j.physletb.2024.139117",
    journal = "Phys. Lett. B",
    volume = "859",
    pages = "139117",
    year = "2024"
}

@article{BZ:18,
    author = "Bondarenko, S. and Zubkov, M. A.",
    title = "{The dimensionally reduced description of the high energy scattering and the effective action for the reggeized gluons}",
    eprint = "1801.08066",
    archivePrefix = "arXiv",
    primaryClass = "hep-ph",
    doi = "10.1140/epjc/s10052-018-6089-1",
    journal = "Eur. Phys. J. C",
    volume = "78",
    number = "8",
    pages = "617",
    year = "2018"
}

@article{MN:19,
    author = "Nefedov, Maxim A.",
    title = "{Computing one-loop corrections to effective vertices with two scales in the EFT for Multi-Regge processes in QCD}",
    eprint = "1902.11030",
    archivePrefix = "arXiv",
    primaryClass = "hep-ph",
    doi = "10.1016/j.nuclphysb.2019.114715",
    journal = "Nucl. Phys. B",
    volume = "946",
    pages = "114715",
    year = "2019"
}

\newpage \ \newpage

\appendix

\begin{widetext}

\section{NLL Green's function}\label{ap:GF}

Several plots of the LL and NLL Green's function~(\ref{eq:Grhs3}) are shown in Fig.~\ref{figA:2}.
At relatively small $\bar Y \propto 0.5$, the behavior of the function w.r.t. $L = \ln |{\bf l}_{\perp 1}| / |{\bf l}_{\perp 2}|$ (Fig.~\ref{figA:2}~(a)--(c)) and $\Delta\psi$ (Fig.~\ref{figA:2}~(d)--(f)) being the azimuthal angle between ${\bf l}_{T 1, 2}$, approximates the initial condition~(\ref{eq:G}) and with rapidity increasing up to $\bar Y \propto 1.5$, the peak spreads.
The CI NLL function does not show any unphysical oscillations or negative values, as evident from Fig.~\ref{figA:2}.
The effect of the CI is numerically small, consistently with observations of Refs.~\cite{Szymanowski:10,Szymanowski:13,Szymanowski:14}.
The NLL function grows slower with the increase of $\bar Y$ than the LL one due to the smaller intercept value  (Fig.~\ref{figA:2}~(g)--(i)).
Finally, we verify the RG-invariance of the Green's function at the NLO accuracy in the region $\bar Y \propto 1$ (Fig.~\ref{figA:2}~(j)--(l)).
In this way, the implementation of the NLO basis of the eigenfunctions together with the CI characteristic function provides the RG-invariant NLL Green's function without any pathologies.

\begin{figure*}[ht]
\centering
    \includegraphics[scale=1.]{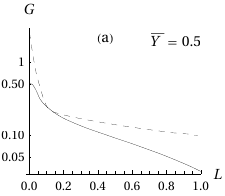}
    \includegraphics[scale=1.]{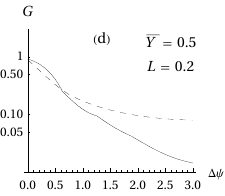}
    \includegraphics[scale=1.]{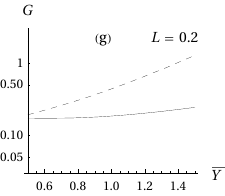}
    \includegraphics[scale=1.]{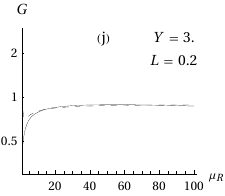}
    \\
    \includegraphics[scale=1.]{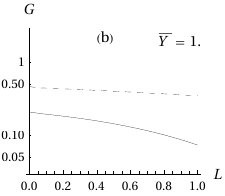}
    \includegraphics[scale=1.]{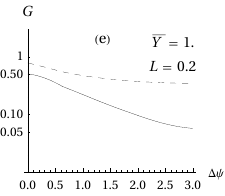}
    \includegraphics[scale=1.]{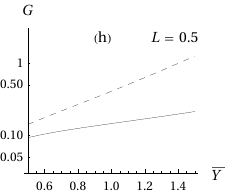}
    \includegraphics[scale=1.]{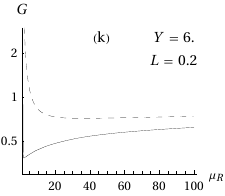}
    \\
    \includegraphics[scale=1.]{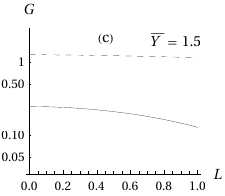}
    \includegraphics[scale=1.]{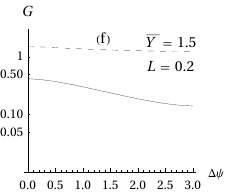}
    \includegraphics[scale=1.]{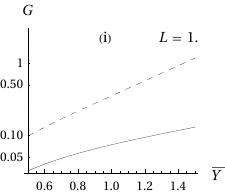}
    \includegraphics[scale=1.]{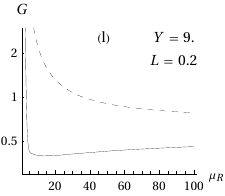}
\caption{\raggedright The plots of the LL and CI NLL Green's function.
In all plots except the ones showing the $\Delta\psi$-dependence, the function is averaged over $\Delta\psi$ is plotted.
The value $\mu_R = 10$ GeV is fixed, except the case of the dependence on it.}\label{figA:2} 
\end{figure*}

\end{widetext}

\end{document}